%% file: p-v2.tex
\documentclass[fleqn,12pt,a4paper]{article}
\usepackage{times}
\usepackage{colordvi}
\usepackage{graphics}
\usepackage{makeidx}
\usepackage{epsf}
\usepackage{here}
\usepackage{feynmp}
\usepackage{cite}
\usepackage{amsmath}
\usepackage{pstricks}
\usepackage{pst-node}
\usepackage{colordvi}
\usepackage{a4wide}
\usepackage{bbold}

\numberwithin{equation}{section}
\numberwithin{figure}{section}
\numberwithin{table}{section}

\usepackage{epsfig}
\usepackage{rotating}

\def \ps {p\hspace{-0.43em}/}

\def \eps {\epsilon}
\def \ep  {\epsilon}
\def \Mcal{\mathcal{M}}

\def \litwo {{\rm{Li_2}}}
\def \litr {{\rm{Li_3}}}

\newcommand{\be}{\begin{equation}}
\newcommand{\ee}{\end{equation}}
\newcommand{\ba}{\begin{eqnarray}}
\newcommand{\ea}{\end{eqnarray}}
\newcommand{\bea}{\begin{eqnarray}}
\newcommand{\eea}{\end{eqnarray}}
\newcommand{\non}{\nonumber}
\newcommand{\nl}{\nonumber \\}
\newcommand{\Fpm}[2]{\Fs{#1}{#2}{1-\frac{p^2}{4m^2}}}
\newcommand{\Ffp}[2]{\Fs{#1}{#2}{\frac{p^2}{4m^2}}}
\newcommand{\Fh}[2]{\,{}_#1F_#2}
\newcommand{\Fs}[3]{\!\!\left[\begin{array}{c}#1\,;\\#2\,;\end{array}#3\right]}

\newcommand{\Bx}{ {\rm Box} }
\newcommand{\VVtt}{\tt V3l1m}

\newcommand{\nn}{\nonumber}
\newcommand{\me}{m}
\newcommand{\rd}{{d}}
\newcommand{\Ao}{A_0}
\newcommand{\Bs}{B_0}
\newcommand{\Bt}{B_t}
\newcommand{\Co}{C_1}
\newcommand{\Cf}{C_4}
\newcommand{\Do}{D_0}

\begin{document}
\allowdisplaybreaks
\input macro
\renewcommand{\thefootnote}{\fnsymbol{footnote}}
\ni
DESY 06--045
\\
BI--TP 2006/21
\\
Freiburg--THEP 06/07
\\
SFB/CPP--06--27

\vspace{1cm}

\begin{center}
{\LARGE \bf
First order radiative corrections
\\[3mm]
to
Bhabha scattering in $d$ dimensions
\footnote{
Work supported in part by the European Community's Human Potential
Programme under contract HPRN-CT-2000-00149 `Physics at Colliders'
 and by Sonderforschungsbereich/Transregio 9 of DFG
`Computergest{\"u}tzte Theoretische Teilchenphysik'.
}}
\\
\vspace{1.5cm}
{
{\Large J. Fleischer}${}^{1}$},~~
{\Large J. Gluza}${}^{2}$,~~
{\Large A. Lorca}${}^{3,4}$~~ and
~~{\Large T. Riemann}${}^{4}$
\footnote{E-mails:~%
fleischer@physik.uni-bielefeld.de, Janusz.Gluza@desy.de,
Alejandro.Lorca@physik.uni-freiburg.de, Tord.Riemann@desy.de
} \\
\vspace{1cm}

{
{${}^{1}$~Fakult\"at f\"ur Physik, Universit\"at Bielefeld, Universit\"atsstr. 25,  33615
Bielefeld, Germany\\ }
\smallskip
{${}^{2}$~Department of Field Theory and Particle Physics,
    Institute of Physics, University of Silesia, Uniwersytecka 4, 40007 Katowice, Poland\\ }
\smallskip
{${}^{3}$~Physikalisches Institut,  Albert-Ludwigs-Universit\"at Freiburg, Hermann-Herder-Str. 3,
79104 Freiburg, Germany\\}
\smallskip
{${}^{4}$~Deutsches Elektronen-Synchrotron, DESY, Platanenallee
  6, 15738 Zeuthen, Germany \\ }
}
\end{center}

\vspace{1cm}

\begin{center}
{\Large \bf
{Abstract}}
\end{center}
The luminosity measurement at the projected International Linear $e^+e^-$ Collider ILC
is planned to be performed with forward Bhabha scattering with an accuracy of the
order of $10^{-4}$.
A theoretical prediction of the differential cross-section has to include one-loop weak corrections, with leading higher order terms, and the complete two-loop QED corrections.
Here, we present the weak part and the virtual one-loop photonic corrections.
For the photonic corrections, the expansions in $\epsilon = (4-d)/2$
are derived with inclusion of the terms of order $\epsilon$ in order to match the two-loop accuracy.
For the photonic box master integral in $d$ dimensions we compare several different methods of evaluation.

\setcounter{footnote}{0}
\renewcommand{\thefootnote}{\arabic{footnote}}

\thispagestyle{empty}

\section{Introduction}
Bhabha scattering
\be \label{eq-1}
e^-(p_1)+e^+(p_4) \to e^-(-p_2)+e^+(-p_3)
\ee
was one of the first processes calculated in quantum
theory \cite{Bhabha:1936xx}.
The complete virtual electroweak one-loop corrections have been first
calculated in \cite{Consoli:1979xw}, later also in
\cite{ Bohm:1984yt,%
Tobimatsu:1985pp,%
Bohm:1986fg,%
Kuroda:1987yi,%
Bardin:1991xe,%
Beenakker:1991mb,%
Montagna:1993py,%
Field:1995dk,%
Beenakker:1998fi%
}.
By now, Bhabha scattering may also be calculated with automated tools for
the evaluation of Feynman diagrams and cross-sections as e.g.
{\tt Feynarts} \cite{Kublbeck:1990xc,Hahn:2000kx},
{\tt grace} \cite{Belanger:2003sd} and
{\tt aITALC} \cite{Lorca:2004fg}.
The electroweak corrections have to be considered together with hard
bremsstrahlung corrections, which usually are calculated by Monte
Carlo programs; see
\cite{Jadach:1996is,Melles:1997qa,Arbuzov:1995qd,Arbuzov:1996jj,Arbuzov:1999db}
and references therein.
Dedicated studies for experimentation at LEP may be found in
\cite{Kobel:2000aw,Jadach:2003zr} and references therein.

The preparation of the $e^+e^-$ linear collider project ILC (formerly
also TESLA \cite{Aguilar-Saavedra:2001rg}, and corresponding projects
of other regions) triggered again some interest in both wide angle and
small angle Bhabha scattering.
The latter might allow to determine the luminosity with an
unprecedented accuracy of $10^{-4}$.
For this, one needs theoretical predictions beyond one-loop accuracy in
the extreme forward scattering region where the cross-section peaks
due to the kinematical singularity of the photon propagators, while
the pure weak corrections might be sufficient in one-loop approximation
(with leading higher order terms a la \cite{Bardin:1991xe}).
If a so-called Giga-Z option will be realized, high Bhabha event rates
are to be expected in the $Z$ resonance region also for larger
scattering angles.

We write the matrix element squared
\begin{eqnarray}
|\Mcal|^2&=&\left(\Mcal^{(0)}+\Mcal^{(1)}+\ldots \right)^*
\left(\Mcal^{(0)}+\Mcal^{(1)} +\ldots\right) + \left( \Mcal_\gamma^{(0)}+ \ldots \right)^* \left( \Mcal_\gamma^{(0)}+ \ldots \right) + \ldots
\nonumber\\
&=& \underbrace{{|\Mcal^{(0)}|}^2}_{\mathcal{O}(\alpha^2)}+ \underbrace{2\Re
({\Mcal^{(0)}}^*\Mcal^{(1)} ) + |\Mcal_\gamma^{(0)}|^2}_{\mathcal{O}(\alpha^3)} \nonumber\\
&&+\underbrace{{|\Mcal^{(1)}|}^2+2\Re
({\Mcal^{(0)}}^*\Mcal^{(2)} ) + 2\Re ({\Mcal_\gamma^{(0)}}^*\Mcal_\gamma^{(1)} ) + |\Mcal_{\gamma\gamma}^{(0)}|^2}_{\mathcal{O}(\alpha^4)} + \mathcal{O}(\alpha^5),
\label{orderexpansion}
\end{eqnarray}
where $\Mcal^{(i)}$ is the contribution to the $i$-loop order and the
subscripts $\gamma$ and $\gamma \gamma$ indicate the emission of one
or two photons.

The QED contributions dominate by far and two-loop corrections are also needed.
Several projects to determine them in a systematic way are underway (see
\cite{Bern:2000ie,Bonciani:2004gi,Czakon:2004tg,Czakon:2004wm,Penin:2005kf,Bonciani:2005im}
and references therein).
A program with two-loop accuracy has to include also the complete
one-loop matrix elements squared,
often regulated by an expansion in
$\epsilon = (d-4)/2$, with a careful treatment of the resulting finite
terms in $\epsilon$.\footnote{A similar program was performed in \cite{Korner:2004rr,Korner:2005rg}.}
For this, one may express the Feynman diagrams by scalar master
integrals, which then have to be known up to some positive order in $\epsilon$.
Thus, one has to go beyond the usual technical demands of a pure one-loop
calculation.

In this article, we
give a concise description of our approach to the one-loop
contributions for a two-loop calculation of massive Bhabha scattering.
Introductory, we present in Section
\ref{electroweak} electroweak predictions which were obtained for the ILC
study \cite{Aguilar-Saavedra:2001rg,Fleischer:2004ah,Lorca:2004dk}.
The expressions for the pure QED corrections up to
order $\epsilon$ in terms of a few scalar master integrals are derived
in Section \ref{qedeps}.
Here we retain the exact dependences on the electron mass.
The scalar master integrals are discussed in Section \ref{masters}.
For the box master integral we compare several, quite different expressions, which are derived with the aid of a difference equation,  a system of differential equations, and the Mellin-Barnes technique, respectively.
We close with a short Summary.

\section{\label{electroweak}Electroweak one-loop corrections}
One-loop corrections are the virtual part of the $\mathcal{O}(\alpha^3)$ terms in (\ref{orderexpansion}).
The calculation of electroweak corrections to Bhabha scattering with the automated tool {\tt aITALC} has been described on several occasions
\cite{Fleischer:2004ah,Lorca:2004dk,Gluza:2004tq,Lorca:2004fg,Lorca:2005yp}.
{\tt aITALC} \cite{Lorca:2004fg} uses the packages {\tt DIANA} v.2.35/{\tt{QGRAF}} 2 \cite{Tentyukov:1999is,Nogueira:1993ex} for the creation of the one-loop matrix elements, {\tt FORM} 3.1 \cite{Vermaseren:2000nd} for their expressions in terms of scalar s, and {\tt LoopTools} 2.1/{\tt{FF}} \cite{Hahn:1998yk,vanOldenborgh:1991yc} for the numerical evaluation, including also soft bremsstrahlung.
In one respect we had to go beyond {\tt LoopTools} 2.1: In order to evaluate cross-sections in the neighbourhood of the $Z$ resonance peak, one has to use Breit-Wigner propagators in the $s$-channel, replacing $m_Z^2$ by $m_0^2 =  m_Z^2 - i m_Z \Gamma_Z$.
Accordingly, the
$\gamma Z$ box function $D_0(t,s,m_0) =  D_0(m^2, m^2, m^2, m^2, t, s, \lambda^2, m^2, m_0^2,m^2)$, if used with $(\gamma, Z)$ in the $s$-channel, has been modified as in
\cite{Beenakker:1990jr}:
\begin{eqnarray}
D_0(t,s,m_0)&=&\frac{x_s}{m_1 m_4 (t-m_0^2)(1-x_s^2)}\Bigg\{
\nonumber\\
&&+2\log(x_s) \left( \log(1-x_s^2) -\log{\frac{m_0 \lambda}{m_0^2-t-i\epsilon}} \right)
\nonumber\\
&&+\frac{\pi^2}{2} +\mathrm{Li}(x_s^2) + \log^2(x_2) + \log^2(x_3)
\nonumber\\
&&-\sum_{\sigma,\rho=\pm1} \left( \mathrm{Li}(x_s x_2^\rho x_3^\sigma) +\left( \log{x_s} + \log{x_2^\rho} + \log{x_3^\sigma}\right)\log(1-x_s x_3^\sigma x_2^\rho) \right) \Bigg\}
\nonumber\\
\eea
The diagram is shown in Figure \ref{ddiagram}.

\begin{figure}[b]
\begin{center}
\begin{fmffile}{fun4}
        \begin{fmfgraph*}(100,100)
        \fmfpen{thin}
        \fmfleft{a4,a1}
        \fmfright{a3,a2}
        \fmf{fermion,label=$p_1$}{a1,i1}
        \fmf{fermion,label=$p_2$}{a2,i2}
        \fmf{fermion,label=$p_3$}{a3,i3}
        \fmf{fermion,label=$p_4$}{a4,i4}
        \fmf{fermion,tension=0.4,label=$m_1$,label.side=left}{i1,i2}
        \fmf{fermion,tension=0.7,label=$m_0$,label.side=left}{i2,i3}
        \fmf{fermion,tension=0.4,label=$m_4$,label.side=left}{i3,i4}
        \fmf{fermion,tension=0.7,label=$\lambda$,label.side=left}{i4,i1}
        \fmfdot{i1,i2,i3,i4}
        \end{fmfgraph*}
      \end{fmffile}\end{center}
\caption{
Four-point function with complex mass $m_0$ and photon mass regulator $\lambda$.
}
\label{ddiagram}
\end{figure}
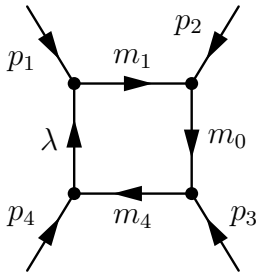
We use the {\tt LoopTools} conventions $s=(p_1+p_4)^2, t=(p_1+p_2)^2, p_i^2=m_i^2=m^2, \lambda=m_{\gamma}$.
with the following definitions for the $x$ variables
\bea
x_s &\equiv& -K(s+i\epsilon,m_1,m_4),\\
x_2 &\equiv& -K(m_2^2,m_1,m_0),\\
x_3 &\equiv& -K(m_3^2,m_4,m_0),
\end{eqnarray}
and the definition of the $K$-function (with one of the arguments being complex)
\begin{equation}
K(z,m,m') \equiv \left\{
\begin{array}{lr}
{  \frac{ 1-\sqrt{1-\frac{4mm'}{z-(m-m')^2}}}{1+\sqrt{1-\frac{4mm'}{z-(m-m')^2}}} }& z\ne (m-m')^2,\\
-1 &  z = (m-m')^2.
\end{array}
\right.
\end{equation}
The photon mass is $\lambda$ in {\tt LoopTools}.
The use of this scalar box function is not only necessary in order to regulate the $\gamma Z$ box contribution, but also for a proper compensation of the corresponding soft photon infra-red divergencies which are proportional to the Born cross-section with a Breit-Wigner $Z$ propagator.
The implementation is done in the file {\tt fortran/src/d0wdd0.F}  of the {\tt aITALC} package.

\def \epz#1{\cdot{} 10^{#1}}
\def\ct{\cos{\theta}}
\def\Oa{\mathcal{O}(\alpha)}
\begin{table}[t]
\begin{center}
$$
\begin{array}{lll}
\hline
\cos{\theta} & \textrm{Born EWSM} & \mathcal{O}(\alpha) \textrm{ EWSM}\\
\hline
 -0.9 & 0.12201 \epz{4} & 0.11767 \epz{4}\\
 -0.7 & 0.10099 \epz{4} & 0.95012 \epz{3}\\
 -0.5 & 0.85685 \epz{3} & 0.79246 \epz{3}\\
 \phantom{+}0\phantom{.0} & 0.73164 \epz{3} & 0.64561 \epz{3}\\
 +0.5 & 0.10701 \epz{4} & 0.91360 \epz{3}\\
 +0.7 & 0.16162 \epz{4} & 0.13917 \epz{4}\\
 +0.9 & 0.70112 \epz{4} & 0.63472 \epz{4}\\
 +0.99 & 0.62198 \epz{6} & 0.57186 \epz{6}\\
 +0.999 & 0.62612 \epz{8} & 0.57540 \epz{8}\\
 +0.9999 & 0.62666 \epz{10} & 0.57822 \epz{10}\\
\hline
\end{array}
$$
\caption{Differential cross-sections in pbarn for Bhabha scattering at
  $\sqrt{s}=m_Z$.
  Born contribution and the $\mathcal{O}(\alpha)$
  correction are shown; the maximum soft-photon energy is $\sqrt{s}/10$.}
\label{table-bhabha-mZ}
\end{center}
\end{table}

\begin{table}[bht]
\begin{center}
$$
\begin{array}{ccllrlr}
\hline
\textrm{rad}&\cos{\theta}&\textrm{Born EWSM}&\multicolumn{2}{c}{\Oa
  \textrm{ EWSM}}&\multicolumn{2}{c}{\Oa \textrm{ QED }N_f=9}\\
\hline
2.691&-0.9\phantom{000}&2.16999\epz{-1}&1.93445\epz{-1}&-10.85\%&4.69800\epz{-1}&116.50\%\\
2.346&-0.7\phantom{000}&2.30098\epz{-1}&2.08843\epz{-1}&-9.24\%&5.03879\epz{-1}&118.98\%\\
2.094&-0.5\phantom{000}&2.61360\epz{-1}&2.38707\epz{-1}&-8.67\%&5.66238\epz{-1}&116.65\%\\
1.571&\phantom{+}0\phantom{.0000}&5.98142\epz{-1}&5.46677\epz{-1}&-8.60\%&1.09322\epz{0}&82.77\%\\
1.047&+0.5\phantom{000}&4.21273\epz{0}&3.81301\epz{0}&-9.49\%&5.13530\epz{0}&21.90\%\\
0.795&+0.7\phantom{000}&1.58240\epz{1}&1.43357\epz{1}&-9.41\%&1.64548\epz{1}&3.99\%\\
0.451&+0.9\phantom{000}&1.89160\epz{2}&1.72928\epz{2}&-8.58\%&1.76464\epz{2}&-6.71\%\\
0.142&+0.99\phantom{00}&2.06556\epz{4}&1.90607\epz{4}&-7.72\%&1.91774\epz{4}&-7.16\%\\
0.045&+0.999\phantom{0}&2.08236\epz{6}&1.91624\epz{6}&-7.98\%&1.92546\epz{6}&-7.53\%\\
0.014&+0.9999&2.08429\epz{8}&1.91402\epz{8}&-8.17\%&1.92270\epz{8}&-7.75\%\\
\hline
\end{array}
$$
\caption{Differential cross-sections in pbarn for Bhabha scattering at
  $\sqrt{s}=500$.
  Born contribution, the $\mathcal{O}(\alpha)$
  correction, and also a QED prediction are shown; the maximum soft-photon energy  is $\sqrt{s}/10$.
}
\label{table-bhabha-500}
\end{center}
\end{table}

In Tables \ref{table-bhabha-mZ} and \ref{table-bhabha-500} we provide numerical sample outputs at typical energies for several scattering angles.
The input quantities as well as the treatment of soft photons are exactly the same as in \cite{Fleischer:2003kk}.
The cross-section peak in the forward direction, due to the photon exchange in the $t$-channel.
In this kinematic region, the pure photonic corrections will be dominating and we have to treat them with higher accuracy than the rest of the electroweak corrections.
For the one-loop corrections, this means a determination of $|\Mcal^{(1)}|^2$
as part of the $\mathcal{O}(\alpha^4)$ terms  in (\ref{orderexpansion}).
Here one needs the QED one-loop functions including terms of order $\eps$ because their interference with other terms of order $1/\eps$ contributes to the finite cross-section.
This will be the main concern of the rest of this article.

\allowdisplaybreaks
\section{\label{qedeps}The massive QED cross-section in $d$ dimensions}
\begin{figure}[bht]
  \begin{center}
\includegraphics[scale=0.75]{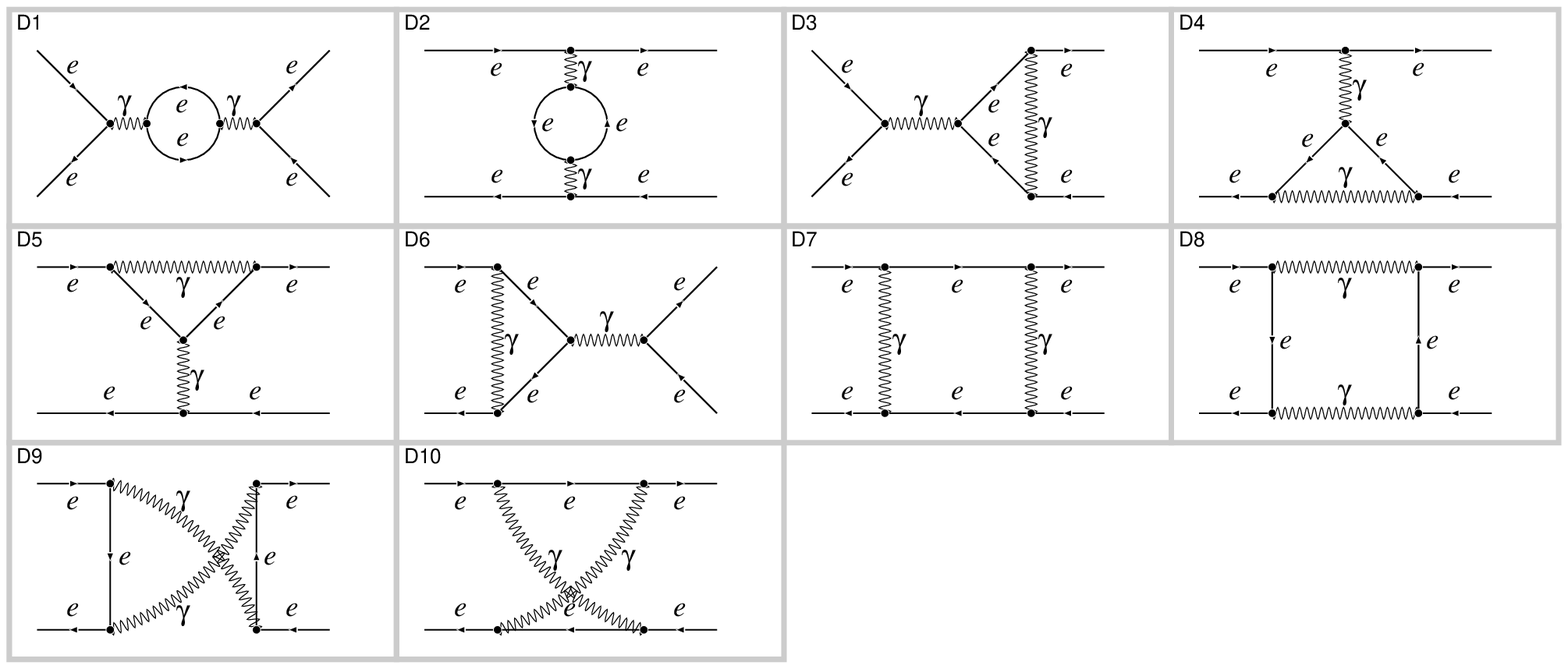}
    \caption{One-loop diagrams for the process $e^+ e^- \to e^+ e^-$.}
    \label{fig:1loop}
  \end{center}
\end{figure}

 The ten diagrams of Fig.\ (\ref{fig:1loop}) are the one-loop  contributions
in pure QED.

We decompose the full one-loop matrix element as follows:
\bea
\label{defmi}
 {\cal M}_1 & = &  \left[ \,   \gamma_{\mu}   \,  \otimes   \,   \gamma_{\mu}    \,   \right]   \, F_1,
 \nonumber \\
 {\cal M}_2 & = &  \left[  \, \ps_4    \,  \otimes   \,  \ps_2   \,   \right]   \, F_2   , \nonumber  \\
  {\cal M}_{3} & = &  \left[ \,  \gamma_{\mu}   \,  \gamma_{\nu}  \,  \gamma_{\rho}  \,  \otimes \,  \gamma_{\rho}  \,  \gamma_{\nu}  \,  \gamma_{\mu}     \,  \right]   \, F_{3},
  \nonumber  \\
  {\cal M}_{4} & = &  \left[ \, \gamma_{\mu}   \,  \gamma_{\nu} \, \ps_4
\,  \otimes  \,  \gamma_{\nu}  \,  \gamma_{\mu}  \, \ps_2  \,  \right]   \, F_{4} ,
\nonumber \\
   {\cal M}_5 & = & ( \left[ \, \ps_4    \,  \otimes   \,  1   \,  \right]   +
\left[  \, 1    \,  \otimes   \,  \ps_2   \,   \right] )  \, F_5,
\nonumber \\
  {\cal M}_6 & = & ( \left[ \, \gamma_{\mu}  \, \ps_4     \,  \otimes   \,   \gamma_{\mu}    \,   \right]  +
                   \left[ \, \gamma_{\mu}       \,  \otimes   \,
\gamma_{\mu} \, \ps_2    \,  \right] )  \, F_6,
\nonumber \\
    {\cal M}_7 & = &  \left[  \, 1   \,  \otimes   \,  1  \,  \right]   \, F_7,
    \nonumber \\
  {\cal M}_8 & = &  \left[ \, \gamma_{\mu}  \, \gamma_{\nu}      \,
\otimes   \,  \gamma_{\nu}\,   \gamma_{\mu}     \,  \right]   \, F_8,
\nonumber \\
  {\cal M}_9 & = & ( \left[ \, \gamma_{\mu}  \,  \gamma_{\nu} \, \ps_4     \,  \otimes   \,  \gamma_{\nu} \,  \gamma_{\mu}     \,   \right]   +
   \left[ \, \gamma_{\mu}  \,  \gamma_{\nu}     \,
\otimes  \,  \gamma_{\nu}  \, \gamma_{\mu}  \, \ps_2    \,   \right]  ) \, F_{9}.
\eea
The notation is short-hand for the $s$ and $t$ channel matrix elements:
\bea
\label{mkst}
{\cal M}_k^{(s)} & = &  O_i \otimes O_f F_k(s,t)
\nl &=&
\bar{v}(p_4) O_i u(p_1) \cdot \bar{u}(-p_2) O_f v(-p_3)  F_k(s,t),
\\
\label{mkts}
{\cal M}_k^{(t)} & = &  O_e \otimes O_p \left[- F_k(t,s) \right]
\nl &=&
\bar{v}(p_4) O_e v(-p_3) \cdot \bar{u}(-p_2) O_p  u(p_1)
\left[- F_k(t,s) \right].
\label{Fermi}
\eea
Crossing the diagrams from the $s$-channel to the $t$-channel results in the exchange of $s$ and $t$ and in an
overall sign change due to Fermi statistics; see Equation (\ref{Fermi}).
In general the first six amplitudes are independent while $F_7$ to
$F_9$ can be expressed in terms of them, but in slightly different
ways for the various cases under consideration.
For this reason we list them all.

With the form factors $F_k(s,t)$ one may determine the contributions from $2\Re({\Mcal^{(0)}}^*\Mcal^{(1)} )$ and from  $|\Mcal^{(1)}|^2$ to the differential cross-section
 (\ref{orderexpansion}).
 The interference of  $\Mcal^{(0)}$ and $\Mcal^{(1)}$ yields e.g.:
\begin{equation}
\frac{d \sigma}{d cos \theta} = \frac{\pi \alpha^2} { 2 s}
          \sum_{j=1}^{9} \left[  B_j(s,t) F_j(s,t) +
                                 B_j(t,s) F_j(t,s) \right],
\label{crossf}
\end{equation}
with
\begin{eqnarray}
B_1 =&&s   \left\{ 4(1+v)^2+2(1-v+\frac{1}{v})-4v z^2+(1-\frac{1}{2v})z^4) \right\}  , \nonumber \\
B_2 =&&s^2 \left\{ -(1+v)[(1+v)^2+v^2]+\frac{1}{2}[(1+v)^2+v(3+5v)-\frac{1}{2 v}]z^2-\right. \nonumber \\
     &&    \left. \frac{3}{4} (2v-\frac{1}{2 v})z^4+\frac{1}{4}(1-\frac{1}{2 v})z^6 \right\}, \nonumber \\
B_3 =&&s   \left\{ 8(3+3v+5v^2)+\frac{8}{v}+8(3-5v+\frac{3}{v})z^2+2(5-\frac{7}{v})z^4\right\} , \nonumber \\
B_4 =&&s^2 \left\{-4(1+v)v^2-(2-6v^2+\frac{1}{v})z^2+(3-4v+\frac{3}{v})z^4     +(1-\frac{5}{4 v})z^6\right\}      ,
\nonumber \\
B_5 =&&{m} s\left\{-2(6v+4v^2-\frac{1}{v})+(2+8v-\frac{3}{v})z^2-(2-\frac{1}{v})z^4\right\}  ,
\nonumber \\
B_6 =&&{m} s\left\{-4(3+\frac{1}{v})+2(1+2v)z^2-(2-\frac{1}{v})z^4\right\} ,\nonumber \\
B_7 =&&s\left\{-\frac{1}{v}-2[1+v-\frac{3}{4 v}] z^2+(1-\frac{1}{2v})z^4\right\} ,\nonumber \\
B_8 =&&s \left\{-\frac{4}{v}+4(1-2v+\frac{3}{v})z^2+(4-\frac{5}{v})z^4  \right\} ,\nonumber \\
B_9 =&&{m} s\left\{-8(3v+4v^2-\frac{1}{v})-4(3-8v+\frac{6}{v})z^2-2(4-\frac{5}{v})z^4\right\},
\label{Bs}
\end{eqnarray}
and
\begin{eqnarray}
v &=& \frac{t}{s},
\\
z &=&\frac{4 m^2}{s}  .
\label{Abbrev0}
\end{eqnarray}
With the same formula (\ref{crossf}), the Born cross-section, arising from $ {{|\Mcal^{(0)}|}^2}$,
is obtained with:
\begin{eqnarray}
F_1^{Born}(s,t)&=&\frac{1}{s},
\\
F_j^{Born}(s,t)&=&0~~~{\mathrm for}~~j~> ~1.
\label{Bornamps}
\end{eqnarray}
The contributions from  $|\Mcal^{(1)}|^2$ to the cross-section are rather lengthy and not shown here explicitely; they will be provided on the webpage \cite{webbhabha1}.
There we give also the expressions for the corresponding interferences in $d$ dimensions.

Before determining the form factors $F_k(s,t)$, we discuss now the various contributions.
As mentioned we may restrict ourselves to the $s$-channel diagrams D1, D3, D6, D8, D9:
\bea
F_j(s,t) &=& \frac{2e^2}{(4\pi)^{d/2}}  \left( F_j^{self} + F_j^{vert} +  F_j^{box} \right), ~~~j=1,\ldots , 9.
\label{d13689}
\eea
The self-energy contributes to
$F_1$ only:
\bea
F_1^{self} &=& F_1^{D1}.
\eea
In a theory with several fermion flavors (with different masses $m_f$), one has to sum this term over all flavors.
The vertices contribute to $F_1$ and  $F_5$:
\bea
F_1^{vert} &=& F_1^{D3} + F_1^{D6},
\label{F1def}
\\
F_5^{vert} &=& F_5^{D3} ~=~ F_5^{D6},
\eea
with $F_1^{D6} = F_1^{D3}$.
The two form factors for ${\cal M}_5$ in (\ref{defmi}) are also equal but contribute to different structures there.
The situation for the box diagrams is a little more involved:
\bea
F_j^{box} &=& c_b\left(F_j^{D8} + F_j^{D9}\right),
\label{Fboxdef}
\eea
and the $c_b$ will be given in (\ref{fboxcb}).
As mentioned only six of the nine form factors are independent.
For the direct box diagram D8 we find the following relations:
\begin{eqnarray}
\label{d789}
F_7^{D8} & = &  4 m^2 \,F_{4}^{D8} + 2 m \,F_5^{D8},
\nonumber \\
F_8^{D8} & = &   m^2 \,F_{4}^{D8},
\nonumber \\
F_9^{D8} & = &   m \,F_{4}^{D8}  - \frac{1}{2} \,  F_6^{D8}  .
\label{threerel}
\end{eqnarray}
There are further relations between the form factors $F_j^{D8}$ from the direct box D8 and the  $F_j^{D9}$ of the crossed box D9:
\begin{eqnarray}
F_1^{D8}  & = & - \, F_1^{D9} + (4 - 6 d) \,F_3^{D9}    ,
\nonumber \\
F_2^{D8}  & = & \, F_2^{D9} -  (4 - 2 d) \,F_4^{D9} ,
 \nonumber \\
F_3^{D8}  & = &  \, F_3^{D9} ,
 \nonumber  \\
F_4^{D8}& = &  - \,F_4^{D9},
 \nonumber \\
F_5^{D8} & = & - 2 d m \,F_4^{D9} - \, F_5^{D9} + d \, F_6^{D9}  ,
\nonumber  \\
F_6^{D8} &= &  F_6^{D9} - 4 m \, F_4^{D9},  \nonumber  \\
F_7^{D8}  & = &  \, F_7^{D9} - (4 - 2 d) \,F_8^{D9} ,
\nonumber  \\
F_8^{D8}& = &  - F_8^{D9} ,
\nonumber \\
F_9^{D8} & = &  \, F_9^{D9}.
\label{crossrel}
\end{eqnarray}
Inverting relations (\ref{crossrel}), diagram D9 is obtained from D8 by exchanging $t$ and $u$.
As mentioned, the $t$-channel box
D7 may be obtained from D8 by simply exchanging $t$ and $s$ (and an overall sign).
Subsequently, diagram D10 results from D7 by inverting again (\ref{crossrel}) and exchanging now
$s$ and  $u$.
As a consistency check,  one can additionally obtain D10 from D9 by
$s,t$ crossing.
The inversion of the first six relations of (\ref{crossrel}) yields $F_1^{D9}$ to  $F_6^{D9}$:
\begin{eqnarray}
F_1^{D9}  & = & - \, F_1^{D8} + (4 - 6 d) \,F_3^{D8} ,
\nonumber \\
F_2^{D9}  & = & \, F_2^{D8} - \, (4 - 2 d)\,  F_4^{D8} ,
 \nonumber \\
F_3^{D9}  & = &  \, F_3^{D8} ,
 \nonumber  \\
F_4^{D9}& = &  - F_4^{D8},
 \nonumber \\
F_5^{D9} & = & - \, F_5^{D8} - 2 d \, F_9^{D8} ,
 \nonumber  \\
F_6^{D9} &= & - F_6^{D8} - 4 \, F_9^{D8}.
\end{eqnarray}
In a next step, one gets in combination with (\ref{d789}):
\begin{eqnarray}
F_7^{D9} & = & -6 d  m^2 \,F_{4}^{D9} - 2 m \,F_5^{D9} + 2 d m \,F_6^{D9} ,
 \nonumber \\
F_8^{D9} & = &   m^2 \,F_{4}^{D9},
  \nonumber \\
F_9^{D9} & = &   m \,F_{4}^{D9}  - \frac{1}{2} \, F_6^{D9}  .
\label{threecro}
\end{eqnarray}
We see that the relations  for $F_7$ in terms of amplitudes
$F_1$ to $F_6$ are different for diagrams D8 and D9.

What remains now is to determine one form factor for the self-energy, two form factors of the vertex, and six form factors for one of the four box diagrams.
This will be done in two steps. First, we collect the form factor contributions from the Feynman diagrams D1 to D10, and in a second step we have to add up additional contributions $F_j^{a,r}$ arising from counter term insertions into the one-loop diagrams.
The latter are formally of higher order, but it is reasonable to discuss them here.
So, effectively, (\ref{d13689}) has to be replaced by
\bea
F_j(s,t) &=& \frac{2e^2}{(4\pi)^{d/2}}  \left[
 F_j^{self} + F_j^{vert} +  F_j^{box} +
 \frac{\delta m}{m} \left(F_j^{self,r}+ F_j^{vert,r}+  F_j^{box,r}
  \right)\right],
  \nn\\ &&j=1,\ldots , 9.
\label{d13689r}
\eea
Additionally, charge renormalization $\delta e/e$ will give an overall factor, and there are also contributions $F_j^{Z}, F_j^{Z,r}$ from wave function renormalization.
Both will be discussed in Section \ref{aux}.

\subsection{\label{ff}The form factors }
We will use the abbreviations for the five master integrals, used here and in the following for the $s$-channel contributions:
\ba
  \label{eq:abb}
A_0 & = & A_0(m)
\\
B_0 &=& B_0(0,0;s)
\\
B_t & =&B_0(m,m;t)
\\
C_0 &=&C_0(m,0,m;m^2,m^2,s)
\\
C_1 &=&C_0(0,m,0;m^2,m^2,s)
\\
D_0&=&D_0(m,0,m,0;m^2,m^2,m^2,m^2,t,s)
\ea
together with the function
\ba
C_4 &=& C_0(m,0,m;m^2,m^2,t).
 \label{eq:abz}
 \ea
The latter may be expressed by $A_0$ and $B_t$, see (\ref{C4}).
This function contains the infra-red singularities and we decided to keep it explicitely as it is also done in {\tt LoopTools}.
 Further we introduce
\begin{eqnarray}
w &=& \frac{u}{s},
\\
x &=&\frac{1}{1-\frac{4m^2}{s}},
\\
y &=&\frac{1}{1-\frac{4m^2}{t}}.
\label{Abbrev}
\end{eqnarray}
In terms of (\ref{eq:abb}) - (\ref{eq:abz}) the results for the
amplitudes are given in the following.
They may also be obtained in {\tt FORM} format from \cite{webbhabha1}.
The explicit expressions for the master integrals are discussed in Section \ref{masters}.
The form factors from self-energies and vertices are:
\begin{eqnarray}
 F_1^{\rm self} =&&  \,  \left.  A_0 \frac{4} {s^2} [\frac{1}{d-1}-1]
 +\,   B_s \frac{2} {s} [\frac{1}{d-1} (1-z)-1] \right. ,
 \\
 F_1^{\rm vert} =&&
2\, \left(- \frac{A_0}{m^2} \frac{1}{s} [\frac{1-x}{d-3}-x z]+\, B_s \frac{1}{s} [x (1+z)+d-4] \right.
\nonumber\\
&&             \left. -\frac{C_0}{(d-3) s} (1+x) (s-4 m^2)  \right) ,
\\
F_5^{\rm vert} =&&  \,    A_0 \frac{2} {m s^2} x [\frac{2}{d-3}-(d-4)]                               +\, B_s \frac{4 m}{s^2} x [1-(d-4)].
\end{eqnarray}
For the box diagrams:
\begin{eqnarray}
F_j^{box}&=& \frac{1}{(vw)^2}\left( F_j^{D8} + F_j^{D9} \right),
\label{fboxcb}
\end{eqnarray}
and the six independent box form factors for that are:
\begin{eqnarray}
F_1^{D8}&=&v \Bigl(~A_d
                [2 v (v_++2 w+v w x)+2 z w_+ y] -
 \frac{4}{ s}[\Bt (w_+ y+v) z              +
                   \Bs v w (1+v x)   ]
 \nn \\
&& - \frac{\Cf}{\rd-3} [(1-3 z) y z^2-v ((1-6 z) y z     -
            2 (2-3 z) z)-v^2 (4+(3 y-10) z)-4 v^3]
             \nn \\
&& +\Co v [ \frac{1}{\rd-3}\{ 2-z-z^2+v (2 w_+ +z)\}-2 (1-z)-2 v (1-w x) ]
  \nn \\
&& -\Do \frac{s v}{2} \left\{ \frac{1}{\rd-3} [\right.
\left. -2 v v_- v_+ \right.
\left. + z(2-z-z^2)+ v  (1+4 z) z -  5 v^2 z]   \right.
\nn \\
&&  \left.+2 v (1-3 z+2 z^2)+v^2 (4-5 z)+ \right.
 \left. 2 v^3  -  (1-z)^2 z \right\}~\Bigr),
\end{eqnarray}

\begin{eqnarray}
F_2^{D8}&=&\frac{2 v}{s}\Bigl(~A_d
               [w+2 v (1+w x)+w_- y]            -
   \frac{2}{s}[\Bt (2 v+w_- y)            +
                     \Bs (1+2 v x ) w ]
 \nn \\
&& -   \frac{\Cf}{\rd-3}  [y_- w_+ + v(2 v -z) -z^2]
 \nn \\
&&  +  \Co\Bigl\{ \frac{1}{\rd-3} [(1-2 v)(1-z) -2 v^2] -(1+2 v x z) (1-z)
 \nn \\
&&+2 v (v (1-x z)-z) \Bigr\}
-\Do \frac{s}{2} \Bigl\{\frac{1}{\rd-3} [(1-z ) (1+w-2 v z)
\nn \\
&&-2 v^2 (w+z) ]+v (1+2 v (w_+ +z)-2 z^2) \Bigr\} \Bigr),
\end{eqnarray}
\begin{eqnarray}
F_3^{D8}&=&\frac{v^2 w }{2}\Bigl(~ - \frac{\Cf}{\rd-3} w_+ +\frac{1}{\rd-3} \{ \Co -  \Do  \frac{s}{2} w_+ \}~\Bigr) ,\\
\nonumber \\
\nonumber \\
F_4^{D8}&=&\frac{v }{s} \Bigl(~
  -\frac{A_d}{2} [w y_-+y) ] +
   \frac{1}{ s} [\Bt y w_-+ \Bs w ] -   \nn \\
&&    \frac{\Cf}{\rd-3} [v^2 -\frac{v }{2} (y_- +z) - (y_- -z)( w_+ - \frac{z}{2}) ]
  -
   \Co [ \frac{1-z}{2} \frac{\rd-4}{\rd-3}
   \nn \\ &&
   -\frac{w}{\rd-3}  -(1-z)]   +
   \Do \frac{s}{4} [v-\frac{w_+}{\rd-3} (w+w_+ -z)]~\Bigr) , \\
\nonumber \\
\nonumber \\
%
F_5^{D8}&=&\frac{2 v \me}{s}\Bigl(~A_d
               [v+(1+v x) w+(1+2 w)y]             -  \\
 &&  \frac{2}{s}[\Bt(v+(1+2 w) y)          +
                        \Bs (1+v x) w   ]         -  \nn \\
&&  \frac{\Cf}{\rd-3} w_+  (y_- -2 w y-z)  -
   \Co w [\frac{1}{\rd-3} - (1+v x_+)] +
   \Do \frac{s w}{2} [\frac{w_+}{\rd-3}  - v]~\Bigr),
\nonumber \\
%
F_6^{D8}&=&\frac{2 v \me}{s}\Bigl(~A_d [v+2 w y+v w x]-
   \frac{2}{s} [\Bt (v+2 w y) + \Bs v w x   ]
\nn \\
&&
   -
   \frac{\Cf }{\rd-3} (v-z) (2 v + w y)  - \nn \\
&& \Co v [ \frac{1 }{\rd-3} +   x (1+v_+ - 2 z)]  -
    \Do \frac{s v}{2} [\frac{1}{\rd-3} (v-z)-1]\Bigr),
\end{eqnarray}
where we have further introduced
\begin{eqnarray}
A_d&=&-\frac{\Ao}{\me^2}\frac{1}{s}\frac{d-2}{d-3},
\\
 x_{\pm}&=&1{\pm} x,
\\
  y_{\pm}&=&1{\pm} y,
\\
   v_{\pm}&=&1{\pm} v,
\\
 w_{\pm}&=&1{\pm} w.
\end{eqnarray}
The small mass limit is easily obtained by putting $z=0, x = y = 1$.
\subsection{\label{aux}Counter term contributions}
In this section we focus on the contributions originating from
renormalization: the charge counter term, the mass counter term
and the wave function renormalization are given in arbitrary dimension.
Not only their $1/\epsilon$ and constant terms are needed in order to
render the amplitudes from the diagrams in Fig. (\ref{fig:1loop})
finite, but also $O(\epsilon)$ terms combine
with divergent parts of the unrenormalized amplitudes
to give additional finite contributions.
Similarly, of course, in two-loop order the
$O(\epsilon)$ contributions of the diagrams combine
with the ${1}/{\epsilon}$ terms of the counter terms to
give finite contributions.

First we consider the charge counter term.
Each
diagram of Figure (\ref{fig:1loop}) has at its vertices a factor
$e$, the electric charge. Renormalization in two-loops
requires $e$ to be replaced by $e(1 + \delta e/e$, with the charge counter term
\be
\frac{\delta e}{e} =-\frac{e^2}{(4\pi)^{d/2}} \frac{d-2}{3} \frac{A_0(m)}{m^2}.
\ee
While the introduction of the charge counter term results only
in an overall factor, the introduction of the mass counter term
is more complicated.
Every internal electron propagator in Fig.~\ref{fig:1loop}  has to be
replaced by:
\begin{eqnarray}
\frac{1}{p_e^2-m^2(1+\delta m/m)^2} &\simeq&  \frac{1}{p_e^2-m^2} \left(
1+\frac{2 m^2 \delta m/m}{p_e^2-m^2} \right),
\label{eqn:massren}
\end{eqnarray}
with
\begin{eqnarray}
  \frac{ \delta m }{m} = \frac{e^2}{(4\pi)^{d/2}}  \,  \frac{(d-1)(d-2)}{2(d-3)} \, \frac{A_0(m^2)}{m^2},
\end{eqnarray}
the electron mass counter-term.
This means that additional amplitudes $F_j^{Dk,r}$ are obtained from the one-loop diagrams D$k$: All
fermion propagators are replaced according to (\ref{eqn:massren}), but the
higher powers of $\delta m/m$ are dropped.
The contributions from the first powers of $\delta m/m$ lead to  'dotted propagators' with
squared numerators.
The second recursion relation given in (\ref{recur3}) reduces the resulting Feynman integrals with dotted lines to master integrals.

Since the  mass renormalization conterterm $\delta{m}/m$ contains
a $1/{\eps}$ pole,
the one-loop master integrals $A_0$ and $B_0$ etc., resulting from the
diagrams,
are needed to order ${\mathcal O}(\eps)$ in order to ensure that all finite terms are taken into account properly.
The relations between the nine amplitudes of a given dotted diagram fulfil
similar relations as those for the undotted diagrams.
The only {\em differences} are the following:
\begin{eqnarray}
F_8^{D8,r} & = &   m^2 \,F_{4}^{D8,r} + \,F_{3}^{D8,r} ,
\label{dotrel1}
\\
F_7^{{D8},r}  & = &  \, F_7^{{D9},r} - (4 - 2 d) \,F_8^{{D9},r} - 4 \, F_3^{{D9},r} ,
\label{dotrel2}
\\
F_7^{D9,r} & = & -6 d  m^2 \,F_{4}^{D9,r} - 2 m \,F_5^{D9,r} + 2 d m \,F_6^{D9,r} +  ~2 d \,F_3^{D9,r},
\\
F_8^{D9,r} & = &   m^2 \,F_{4}^{D9,r}  - \,F_3^{D9,r}    .
\label{dotrel3}
\end{eqnarray}
All other relations remain unchanged.

Now we present the contributions to the amplitudes
for the dotted diagrams:
\begin{eqnarray}
 F_1^{self,r}  &=& \,  \left.  A_0 \frac{4} {s^2} (d-2) x z    \,  - B_s \frac{2} {s  } z [1-(d-3) x] \right. ,
 \\
 F_1^{vert,r} &=&  2 \Bigl(
 A_0 \frac{2} {s^2} x \left[(\frac{-1}{d-3}+6 \frac{1}{d-5})
    -(d-4) (2   x-1+(d-4))+7-4 x \right]
    \nonumber \\
&& +\,A_0 \frac{2} {m^2 s} [1+3 \frac{1}{d-5}]
  -\,B_s \frac{1} { s}  x z [(d-4) (2 x+(d-4))+2 x]
   \nonumber \\
&&   +  \frac{1}{(d-3) s} x z (s-4 m^2) C_0   \Bigr),
 \\
 F_5^{vert,r}  &=&
 - \frac{A_0}{m} \frac{4} {s^2} x \left[\frac{-1}{d-3} +(d-4)(1+(d-4)(1-\frac{x}{2}))-3+2 x \right]
                            -B_s\frac{4 m}{s^2} x  \nonumber  \\
&&\left[(d-4)(x-(d-4)x z) + 2 x z \right ]-
 \frac{4 m}{(d-3) s^2} x (1+x) (s-4 m^2) C_4  .\nonumber
\\
\end{eqnarray}
For the box diagrams:
\begin{eqnarray}
F_j^{box,r}&=& \frac{1}{(vw)^2}\left( F_j^{D8,r} + F_j^{D9,r} \right),
\end{eqnarray}
and the six independent dotted box form factors for that are:
\begin{eqnarray}
 F_1^{D8,r}&=&
 \frac{\Ao}{\me^2}\frac{1}{s}\left\{ (d-4) \Bigl[2 v^2 (v_+(1+v_+)-x v w + 2 w z)-\right.
        \left.        z v \Bigl(v (2 + v_+) - w y (1+2 w)
\right. \nn \\
&&+z( v  + y) \Bigr)      \left.   - z^2 y^4 w_+\Bigr]+ \frac{6}{d-5} \Bigl[v^2 (v_+(1+v_+) -v w +2 w z)
         + \frac{v z}{2}  (v (2 w - 3 z)
\right. \nn \\
&&   +2y(w_+^2 -w) )  +\frac{v^2 z^2}{2} \Bigr] +     \left. 2 v^2 [v (4 + 2 w_+ + 3 z)+\right.
         \left.       6 (w_+^2-w)-2 x v w]
\right. \nn \\
&&  \left. + \frac{v z}{2} \Bigl[w_+ -8 v (1+v_+ +\frac{3 z}{4})    +y (3+17 w+12 w^2)\Bigr]+ \right. \
        \left.      \frac{z^2}{4} [v (8 v-2-4 y w_-) \right.
\nn \\
&&  - 4 y^4 (2+w)]\Bigr\}
  + \Ao  \frac{2}{s^2 (d-3)}\Bigl\{ v [y_-  +w (1+9 y) + 2  v z - (1-2 w_+ y) z ]
\nn \\
&& +  2 w y^4 z \Bigr\} -  \Bt  \frac{z}{s} \left\{  4 (d-4) [v (v+y w_+ )+\frac{y z}{2}  (v+y w_+ )]   +
                 v \Bigl(v + 4 v (1-z)
\right.
\nn \\
&&  + 5 y w_- -z\Bigr) + z   \left.   \Bigl(v + 2 v^2 -2 y (v w-y) \Bigr)  \right\} - \Bs     \frac{v z (d-3)}{s}
  \Bigl\{ 2 v (w + w_+)
\nn \\
&&+   2 x v w(1+2 v -z) +
2 y (w^2 + w_+^2)\Bigr\}+ \Cf \frac{2}{d-3} \left\{  v \left[ v \left( v z ( w_+ + v -\frac{3 z}{2})\right.  \right.  \right.
   \left. \left.  \left.
   \right.  \right.  \right.
   \nn \\
&& +  [y_- +w (1+9 y)-
\left. \left.  \left.(5-2 y+2 w (2-y)) z] \frac{z}{4}+ \frac{z^3}{2} \right) -
 [y_- +w (1+9 y-2 y^4 )]
 \right. \right. \nn \\
&&
\frac{z^2}{4} + (1-2 w_+ y) \frac{z^3}{4} \Bigr]
\left.  -w y^4 \frac{z^3}{2} \right\}+
   \Co    \frac{v z}{2}    \left\{  (d-4) \Bigl[ w_+ -v (1+2 w   +
    8 w^2-4 v^2)
\right. \nn \\
 && + 4 x v^2 w+  2 y ( w^2 +\left.w_+^2)- z (1-4 v w_+ +2 v^2+2 y ( w^2 + w_+^2) \right.
           \left.    )-2 z^2 v\Bigr]
 + \right.\nn \\
&&-  2 v (w+2 v v_+ -   2 x v w)
 + \left.  z\Bigl(w_+ + (3 v -z) (v+v_+)\Bigr)   \right\}    +
   \Do  v z s  \Bigl\{(d-4)
 \nn \\
&& \Bigl[ v v_+^2  -\frac{z}{2} \Bigl( v (2 - 2 w + 3 v)- y
 ( w^2+w_+^2)\Bigr)+  \frac{z^2}{2}  v \Bigr]-v \Bigl(2+(2 v + w) w_+ +\frac{w}{2} \Bigr)
\nn \\&&+      \frac{z}{2} \Bigl(v (3+2 w)-y ( w^2+w_+^2)\Bigr)  \Bigr\},
\end{eqnarray}\begin{eqnarray}
F_2^{D8,r}&=&
 \frac{ \Ao}{\me^2}\frac{1}{s^2} \left\{(d-4) \left[2 v
\Bigl(2 v (2+w x_-+z) +4 w_+y - w y_-\Bigr)  -2 z
\Bigl(2 v (v+y)
\right.\right. \nn \\
&& +(3+5 w)\left. \left.  y^4\Bigr)\right] - \frac{2}{d-3} v w y_- +
             \frac{24 v}{d-5}  [v + w_+ y]+
            2 v \Bigl[6 v (3+v)
 \right. \nn \\
&&    -(3-10 v+4 v x-15 y)  w   +    \left.   12 y\Bigr]    -2 z \Bigl[4 v^2-2 v w y+(6+9 w) y^4\Bigr]\right\}
  \nn \\
&& + \Ao \frac{8}{s^3 (d-3)} \left\{w y^4+2 v (v+w y+2 y)\right\}
-\Bt       \frac{4}{s^2}\left\{ z (d-4) \Bigl[2 v y+(3+5 w) y^4\Bigr]\right.
            \nn \\
&&+ v w y_- - z \Bigl( 2 v (v+y w_+)-(3+4 w) y^4\Bigr)\Bigr\}
\nn \\ 
&& + \Bs  \frac{4 v}{s^2} \Bigl\{ (d-4) [y_- -z (1+2 x v w)]  +y_-  - z (1+2 x v w)\Bigr\}
\nn \\
&&
  -  \Cf \frac{2}{s (d-3)} \Bigl\{ v^2 w y_-
      v z [ w (1-y y_-)+  2 v (v+y (1+w_+))]
   \nn \\&&
      +  z^2 [w y^4 + 2 v y (2+w)+2 v^2] \Bigr\}
    \nn \\
&&  -   \Co  \frac{2 v }{s} \left\{(d-4) [y_- - z (1- 2 v (w_+ -x w)-y) ] +2 z v (1-w x_+)           \right\}
  \nn \\&&
  +    \Do  v z   \left\{(d-4) [2 v w_+ +y] -y_+ +z\right\},
\nn \\
\end{eqnarray}
\begin{eqnarray}
F_3^{D8,r}&=&
-\Ao     \frac{d-2}{d-3}       \frac{2}{s^2}   v w y-
 \Bt  \frac{1}{s}   v w y z-
 \Cf   \frac{1}{2 (d-3)}         v w (v-z) y z+
 \Co         \frac{1}{2}       v^2 w z- \nn \\
&& \Do          \frac{1}{4} s     v^2 w z,
\end{eqnarray}\begin{eqnarray}
F_4^{D8,r}&=&\frac{\Ao } {{\me}^2} \frac{1}{s^2 } \Bigl\{
\frac{1}{d-3} \left[ -\frac{v}{2} (y_- -z) +(v-\frac{w}{2}  y^3) y z\right]+
   \frac{3 v}{d-5}  \left[y w_+ +v \right]
\nn \\&&
   +  (d-4)\Bigl[\frac{1}{2} v w y_+  + v (v+y)-\frac{1}{2} w_- y^4 z\Bigr]
\nn \\
&&
+
    v \left[(w+ \frac{z}{2}) (1+2 y)+\frac{7}{2} y+3 v-\frac{1}{2}\right]+\frac{1}{2} (w-2) y^4 z\Bigr\}
    \nn \\
&&  +\Bt \frac{1}{s^2} \left\{\Bigl(v (-y_- + (1+2 y) z)-y^4 z\Bigr)-(d-4) y^4 z (1+v_+ -z) \right\}
\nn \\ &&
-    \Bs (d-3) \frac{v}{s^2}\left\{y_- + 2 v x  (v_+ -z) (1-z)-2 v (y+v-z)- z (1-2 y)\right\}
\nn \\ &&
+ \Cf \frac{1}{s} \frac{1}{2(d-3)} \left\{ (v (1+ 2 z)- w y^3 z) y-(1-z) v \right\}(v-z)
\nn \\ &&
 -    \Co \frac{v}{s} \left\{v z-(d-4) \Bigl[(w+\frac{1}{2}) y(1-z) +   \frac{1}{2} (v-w)
    - 
 v z\Bigr] \right\}
\nn \\ &&
+ \Do v z \left\{\frac{y}{4} (1-2 z)+\frac{1-z}{4}+\frac{1}{2} v y_+ -(d-4) [\frac{y}{4} (1-2 z)-\frac{v}{2}
                     y_- ]\right\},
\end{eqnarray}\begin{eqnarray}
F_5^{D8,r}&=&
  \frac{\Ao}{\me} \frac{1}{s^2}  \left\{ (d-4) \Bigl[-2 v (v+w+x v w+y w_+ )+\right.
              \left.  2 z y^4 w\Bigr]-\frac{6 v}{d-5} (y w_+ +v) +\right. \nn \\
&&  \left. 2 v \Bigl[1-3 v-2 w-2 x v w-y (4+3 w)+\frac{y_--z}{d-3} \Bigr]  -\right.
      \left.  2 z (v-2 y^4 w)                            \right\}
       + \nn \\
&&\Bt   \frac{ 4 \me }{s^2}\left\{ v y_- +  z ((d-3) y^4 w-v) \right\}
\nn\\&&
+
  \Bs   \frac{ 4 \me v}{s^2}(d-3) [v-w+y-2 v w x (1-\frac{z}{2})]   + \nn \\
&&\Cf \frac{2 v \me}{s}  \frac{1}{d-3} (v-z) (y_- -z)                   +
  \Co  \frac{ 2 v \me}{s} \left\{(d-4) \Bigl[2 v w-v+w-y+z (v+y + \right.  \nn \\
&& \left. x v w)\Bigr] +
 z x v w \right\}  - \Do 2 v  \me  \left\{\frac{z}{2} (d-4) (v+y)+v v_+ w + \right.
         \left.         \frac{z}{2} [y_- -v w-z ]     \right\},
\nn\\
\end{eqnarray}\begin{eqnarray}
F_6^{D8,r}&=&
  \frac{\Ao}{\me} \frac{1}{s^2 } \left\{ (d-4) \Bigl[2 v (v-x v w+y)-2 z y^4 w_- \Bigr]+ \right.
           \left.     \frac{6}{d-5} [v (v+y w_+)]+\right. \nn \\
&&\left.   2 v \Bigl(3 v-2 x v w+  3 y\Bigr)+2 z y \Bigl(2 v-y (2-w)\Bigr) \right\}                -
   \frac{\Ao}{\me} \frac{2 y}{s^2 (d-3) }   \left\{ v w+z (w y-
\right.\nn \\
&& \left. 2 v)\right\}
 -\Bt  \frac{4 \me }{s^2} \left\{(d-4) z y^4 w_- +y v w-z y (2 v-y) \right\}     -
   \Bs  \frac{4 \me }{s^2} \left\{(d-3) (v (v+y) \right. \nn \\
&& \left. w+z x v^2 w)\right\}              -
\Cf   \frac{2 \me y }{s (d-3)}  \left\{-w y z^2+v (w (v-y_- z)        -
               2 (v-z) z)\right\}                               +\nn \\
&& \Co \frac{ 2  v \me}{s}   \left\{(d-4) [(v+y) w-z (v-w (v x-y))]          +
               v w-z (2 v-x v w)\right\}                         +\nn \\
&& \Do    v \me   \left\{(d-4) [z (v+y w)]+v w (v + w)  + z (v (v_- -2 w)-y w)+z^2 v\right\}.
\end{eqnarray}

Finally we investigate wave function renormalization for the electron self-energy:
\begin{eqnarray}
\Sigma (p)=A(p^2)+B(p^2) ({\slash p}- m).
\end{eqnarray}
The wave function renormalization is given by:
\begin{eqnarray}
Z = 1 + B + 2 m \frac{\partial A}{\partial p^2} |_{p^2=m^2} = 1 + \delta Z,
\end{eqnarray}
and the `undotted' one then reads
\begin{eqnarray}
\delta Z = -\frac{e^2}{(4\pi)^{d/2}} \left\{\frac{d-2}{2}\frac{A_0(m)}{m^2}+4 m^2 DB_0(0,m,m^2) \right\}
\label{undotZ}
\end{eqnarray}
with
\begin{eqnarray}
DB_0(0,m,m^2)=\frac{\partial B_0}{\partial p^2} |_{p^2=m^2}.
\label{AB0}
\end{eqnarray}
The first part in (\ref{undotZ})  contains the
UV divergence and the second  the infrared divergence.
Explicitely (\ref{AB0}) reads
\begin{eqnarray}
DB_0(0,m,m^2)&=&\frac{(d-2)}{(d-3)}\frac{1}{4 m^2}\frac{A_0(m)}{m^2}
\non\\
&=&\frac{1}{(m^2)^{3-\frac{d}{2}}}\frac{\Gamma(3-\frac{d}{2})}{(d-3)(d-4)}
\non\\
&=&\frac{C_0(m,0,m;m^2,m^2,0)}{(d-3)}.
\label{BBAB}
\end{eqnarray}
The UV divergent part of the wave function renormalization
cancels the UV divergence of the vertex, and a remaining IR-singularity will be compensated by soft photon radiation.
It is worth mentioning that due to (\ref{BBAB}) we can also write
\begin{eqnarray}
\delta Z = -\frac{\delta m}{m} .
\end{eqnarray}

We now discuss the dotted diagrams.
They are
UV finite and the divergent contributions to the `dotted` $\delta Z$
come only from the IR divergence.
Therefore we write  the wave function renormalization from the dotted self-energy $\delta Z^{r}$
in terms of $DB_0$ (or $C_0$, respectively):
\begin{eqnarray}
\delta Z^{r} = - 2 \frac{e^2}{(4\pi)^{d/2}} m^2 \frac{\delta m}{m} (d-2) \left[ 6+4(d-4)-(d-4)^2\right] DB_0(0,m,m^2).
\end{eqnarray}

The resulting form factors are:
\begin{eqnarray}
F_j^{Z(,r)}(s,t) &=& \frac{4}{s} \delta Z^{(r)}.
\label{Z}
\end{eqnarray}
The true  one-loop form factors contribute to the interference with Born (as shown here explicitely) as well as to the squared one-loop correction (not shown explicitely), while the dotted form factors contribute only to the former.
\section{\label{masters}The master integrals}
\begin{figure}[!ht]
\begin{center}
  \begin{fmffile}{masters-bhabha}
    \begin{fmfgraph*}(100,100)
      \fmfstraight
      \fmfleft{e1}
      \fmfright{e2}
      \fmf{dashes}{e2,i1}
      \fmf{dashes}{e1,i1}
      \fmf{plain,tension=0.8}{i1,i1}
      \fmfbottom{l1}
      \fmfv{label.angle=90,label={\tt T1l1m}}{l1}
    \end{fmfgraph*}
    \qquad
    \begin{fmfgraph*}(100,100)
      \fmfstraight
      \fmfleft{e1}
      \fmfright{e2}
      \fmf{dashes}{e2,i2}
      \fmf{dashes}{e1,i1}
      \fmf{plain,left,tension=0.4}{i1,i2}
      \fmf{plain,left,tension=0.4}{i2,i1}
      \fmfbottom{l1}
      \fmflabel{Topology 2}{l1}
      \fmfv{label.angle=90,label={\tt SE2l2m}}{l1}
    \end{fmfgraph*}
    \qquad
    \begin{fmfgraph*}(100,100)
      \fmfstraight
      \fmfleft{e1}
      \fmfright{e2}
      \fmf{dashes}{e2,i2}
      \fmf{dashes}{e1,i1}
      \fmf{photon,left,tension=0.4}{i1,i2}
      \fmf{photon,left,tension=0.4}{i2,i1}
      \fmfbottom{l1}
      \fmfv{label.angle=90,label={\tt SE2l0m}}{l1}
    \end{fmfgraph*}
    \newline
    \begin{fmfgraph*}(100,100)
      \fmfstraight
      \fmfleft{e1}
      \fmfright{e3,e2}
      \fmf{plain}{e3,i3}
      \fmf{plain}{e2,i2}
      \fmf{dashes}{e1,i1}
      \fmf{photon,tension=0.4}{i2,i1}
      \fmf{photon,label.side=right,tension=0.4}{i1,i3}
      \fmf{plain,tension=0.3}{i3,i2}
      \fmfbottom{l1}
      \fmflabel{{\tt V3l1m}}{l1}
    \end{fmfgraph*}
    \qquad
    \begin{fmfgraph*}(100,100)
      \fmfstraight
      \fmfleft{e4,e1}
      \fmfright{e3,e2}
      \fmf{plain}{e4,i4}
      \fmf{plain}{e3,i3}
      \fmf{plain}{e2,i2}
      \fmf{plain}{e1,i1}
      \fmf{photon,tension=0.4}{i1,i4}
      \fmf{photon,tension=0.4}{i3,i2}
      \fmf{plain,tension=0.4}{i1,i2}
      \fmf{plain,tension=0.4}{i3,i4}
      \fmfbottom{l1}
      \fmflabel{{\tt B4l2m}}{l1}
    \end{fmfgraph*}
    \qquad
  \end{fmffile}
\end{center}
\vspace*{0.1cm}
\caption{The five one-loop MIs. External solid (dashed) lines describe
on- (off-)shell momenta.}
\label{beta1}
\end{figure}
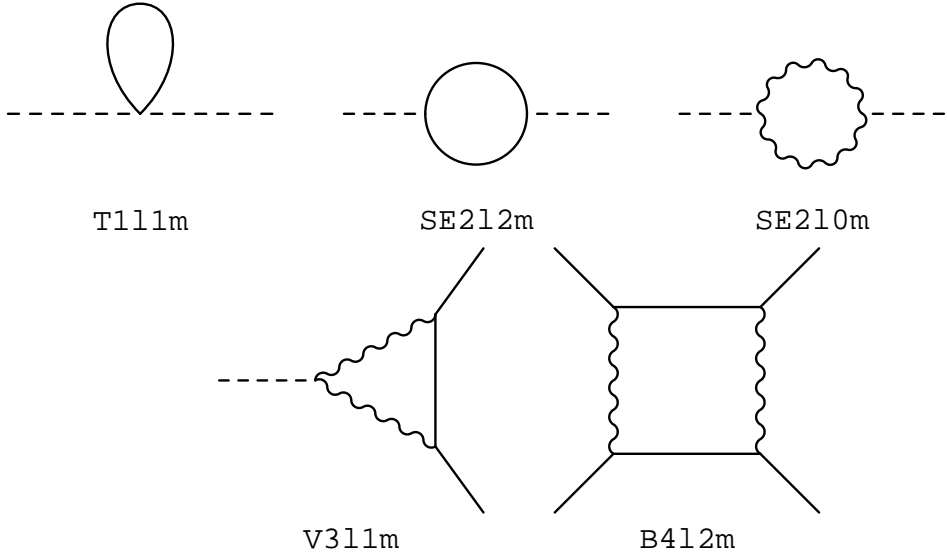

The five master integrals of massive Bhabha scattering are shown in
Figure \ref{beta1}.
We collect here expressions for them valid in $d$ dimensions, but
also the necessary $\epsilon$-expansions.

\subsection{\label{masterA}One-point function}
The simplest master integral is the tadpole:
\footnote{We omit here and in the following the conventional scale
  factor $(4\pi\mu^2)^{\eps}$; the scale
  factor would make the arguments of logarithms dimensionless.}
\ba
A_0(m)
&=&
- \frac{e^{\epsilon \gamma_E}}{i\pi^{d/2}}
\int \frac{d^d k}{k^2-m^2}
\nl
&=&
\Gamma(1-d/2)(m^2)^{\frac{d-2}{2}}
\nl
&=& -{m}^2 \Bigl[ \frac{1}{\eps}+(1-L_m)+
         \frac{\eps}{2} \left(2+\zeta_2-2L_m+ L_m^2\right)
\nl &&
+ \frac{\eps^2}{6}
\left(6 + 3\zeta_2  - 2\zeta_3 - 3(2 + \zeta_2)L_m + 3L_m^2 - L_m^3 \right)
       \Bigr]
+ \ldots ,
\ea
with the abbreviation
\begin{eqnarray}
L_m&=&\ln(m^2).
\end{eqnarray}
Often, shorthand notations with $m=1$ are used, and our tadpole
formula then agrees with {\tt T1l1m} as it is given in the
{\tt Mathematica}
file  {\tt MastersBhabha.m} located at \cite{web-masters:2004nn}:
\begin{eqnarray}
- A_0(1) &=&
 {\tt T1l1m}
\nl
&=&
\frac{1}{\epsilon} +1
+\left(1+ \frac{\zeta_2}{2}\right)  \epsilon
+\left(1+ \frac{\zeta_2}{2} - \frac{\zeta_3}{3}\right)  \epsilon^2
+\ldots
\label{norm}
\end{eqnarray}
\subsection{\label{masterB}Two-point functions}
The two-point functions are
\ba
B_0(m,M;p^2)
&=&
 \frac{e^{\epsilon \gamma_E}}{i\pi^{d/2}}
\int \frac{d^d k}{ (k^2-m^2) [(k+p)^2-M^2] } .
\ea
There are two of them, $B_0(0,0;p^2)$ (coming from the
reduction of box diagrams) and $B_0(m,m;p^2)$.
In $d$ dimensions, they have been determined in \cite{'tHooft:1972fi} and in
\cite{Davydychev:2000na}, correspondingly:
\begin{eqnarray}
B_0(0,0;p^2)
&=&
\frac{e^{\epsilon \gamma_E} \sqrt{\pi} }{(-p^2)^{(2-\frac{d}{2})}}~
\frac{\Gamma \left(2-\frac{d}{2} \right)
      \Gamma \left(\frac{d}{2}-1 \right)}
      {2^{d-3} ~\Gamma \left(\frac{d-1}{2}\right)}
\\
B_0(m,m;p^2)
 &=&e^{\epsilon \gamma_E}(m^2)^{-(2-\frac{d}{2})}~
 \Gamma \left( 2- \frac{d}{2} \right)
 \Fh21\Ffp{1,2-\frac{d}{2}}{\frac{3}{2}} .
\end{eqnarray}
The $\eps$-expansion for $B_0(0,0;p^2)$ is trivial,
\ba
B_0(0,0;p^2)
&=&
\frac{1}{\eps} +2 - \ln(-p^2) +\eps \left[4- \frac{\zeta_2}{2}
-2\ln(-p^2) +\frac{1}{2} \ln^2(-p^2) \right] + \dots,
\ea
and the one for $B_0(m,m;p^2)$ may be determined by using a relation for
contiguous hypergeometric functions
\ba\label{eq-hyp2}
\Fh21\Ffp{1,2-\frac{d}{2}}{\frac32}
&=& \frac{1}{1-2\epsilon} \left\{1-2\epsilon\left(1-\frac{p^2}{4m^2}\right)
\Fh21\Ffp{1,1+\epsilon}{\frac32}\right\} ,
\ea
and then expanding the transformed hypergeometric function
\cite{Davydychev:2000na,Davydychev:2003mv,Kalmykov:2006pu,webpage-kalmykov}.
\footnote{We thank M. Kalmykov for the {\tt FORM}
  code {\tt hypergeometric2F1} for an automatized derivation of the $\eps$-expansion.}
The result is:
\ba
B_0(m,m;p^2)
 &=&
\frac{1}{\eps} + 2-L_m +r \ln(x) + \eps \Bigl[4 + \frac{\zeta_2}{2}
  - 2 L_m + \frac{1}{2} L_m^2  +
r \Bigl( 2 \ln(x)
\nonumber \\
&&- \ln(x) L_m - 2\ln(x)\ln(1 + x) + \frac{1}{2}\ln^2(x) -\zeta_2 -
2\litwo( - x) \Bigr) \Bigr]
\nonumber \\
&& + {\eps}^2 \Bigl[
- 4 L_m + L_m^2 + \zeta(2) - \frac{1}{2} \zeta(2) L_m
 + r    \Bigl(  - 2 \zeta(2) + \zeta(2) L_m \non \\
&&- 2 \zeta(3) - 4 \litwo(-x) + 2 \litwo(-x)
         L_m - 2 \litr(-x) + 4 S_{1,2}( - x) \non \\
&& +      2   \ln(1 + x)  ( \zeta(2) + 2 \litwo(-x) ) +
        2 \ln(x) \ln^2(1 + x) \non \\
&& - ( 4 - 2 L_m ) \ln(x) \ln(1 + x) -
        \ln(x)^2 \ln(1 + x) + \frac{1}{6} \ln(x)^3 +  \non \\
&& ( 1 -\frac{1}{2}  L_m ) \ln(x)^2 +
        ( 4 - 2 L_m + \frac{1}{2} L_m^2 - \frac{1}{2} \zeta(2) ) \ln(x) \Bigr)
 \Bigr] + \cdots ,
\ea
with
\ba
x&=&
\frac{\sqrt{1-\frac{4 m^2}{p^2}}-1}{\sqrt{1-\frac{4 m^2}{p^2}}+1}
~\equiv~ \frac{1-b}{1+b},
\\
b&=&\sqrt{\frac{p^2}{p^2-4 m^2}},
\\
r&=&\frac{1+x}{1-x} ~=~\frac{1}{b} .
\ea
The $\eps$-expansions may also be determined by the method of
differential equations \cite{Kotikov:1991hm,Remiddi:1997ny} and are
then naturally expressed in
terms of Harmonic Polylogarithms
\cite{Remiddi:1999ew,web-masters:2004nn}. With $m=1$ we have \cite{web-masters:2004nn}:
\ba
\label{b000p}
B_0(0,0;p^2)
 &=&
 {\tt SE2l0m(x)}
\nl
&=&
\frac{1}{\eps} +
2 + H[0, x] + 2~H[1, x]
+\eps \Bigl\{
(4 - \zeta_2/2 + 2~H[0, x] + 4~H[1, x]
\nl&&
+ H[0, 0, x] +
  2~(H[0, 1, x] + H[1, 0, x]) + 4~H[1, 1, x] )
\Bigr\} +\ldots
\\
\label{b0mmp}
B_0(m,m;p^2)
 &=&
 {\tt SE2l2m(x)}
\\
&=&
\frac{1}{\eps}
+
2 + \frac{1 + x}{1-x}~H[0, x]
+\eps \Bigl\{
( (-8 + 8~x + \zeta_2 
\nl&&
+ 3~x~\zeta_2 - 4~(1 + x)~H[0, x]
+
  4~(1 + x)~H[-1, 0, x] 
\nl&&
- 2~H[0, 0, x] - 2~x~H[0, 0, x])/(2~(-1 + x))
  )
\Bigr\}
\nl
&&+\eps^2 \Bigl\{
( ((1 + x)~(8 - 16/(1 + x) + 3~\zeta_2 - (2~\zeta_2)/(1 + x) \nl&&
+
  (5~\zeta_3)/3 + (2~\zeta_3)/(3~(1 + x)) - 2~\zeta_2~H[-1, x] \nl&&
+ ((-8 + \zeta_2)~H[0, x])/2 +
  4~H[-1, 0, x] - 2~H[0, 0, x] - 4~H[-1, -1, 0, x] \nl&&
+ 2~H[-1, 0, 0, x] +
  2~H[0, -1, 0, x] - H[0, 0, 0, x]))/(-1 + x) )
\Bigr\}
+ \ldots
\nonumber
\ea
With the {\tt Mathematica}
file  {\tt HPL4.m}., also located at \cite{web-masters:2004nn}, the
corresponding expressions
in terms of polylogarithms may be derived from (\ref{b000p}) and
(\ref{b0mmp}).

\subsection{\label{masterC}Three-point functions}
There are two three-point functions,
$C_0(0,m,0,m^2,m^2,p^2)$
and
$C_0(m,0,m,m^2,m^2,p^2)$, with the definition ($p=p_1+p_2$):
\ba
C_0(m_1,m_2,m_3,m^2,m^2,p^2)
&=& -
 \frac{e^{\epsilon \gamma_E}}{i\pi^{d/2}}
\\
\nonumber
&&
\int \frac{d^d k}{ [k^2-m_1^2] [(k+p_1)^2-m_2^2] [(k+p_1+p_2)^2-m_3^2]  } .
\ea
As shown in (\ref{C4}) $C_0(m,0,m,m^2,m^2,p^2)$ is not a master integral.
The UV-divergences of $A_0$ and $B_0$ in (\ref{C4}) cancel, and the
factor $1/(d-4)$
represents the IR-divergence of this vertex function.\footnote{The
loop  functions $A_0$ and $C_0$ used here deviate by an overall sign
from conventions of e.g. \cite{Hahn:1998yk,looptools}.}
Due to the additional factor of $1/\eps$, we need $A_0$ and $B_0$
up to $O(\eps^2)$ for a $C_0$ of order $\eps$.
As discussed above, a separate control of IR divergences is often
quite helpful in applications; therefore the explicit use of
$C_0(m,0,m;m^2,m^2,s)$ is recommended and we reproduce it here for
completeness (see also Equation 39 of \cite{Davydychev:2000na}):
\begin{eqnarray}
C_0(m,0,m,m^2,m^2,p^2)
&=&
\frac{-1}{(p^2-4 m^2) b}
\Bigl\{ \ln(x) \frac{1}{\eps}
-
\Bigl( \ln(x) [\ln(-p^2)+\ln(-x) ]
\\&&
+2\litwo(-x)-\frac{1}{2}\ln^2(x)+\zeta_2 \Bigr)
+
\eps
\Bigl(\frac{1}{6} \ln^3(x)
+ \Bigl[2 \litwo(-x)
\nonumber \\  &&
-\frac{1}{2}\ln^2(x)+\zeta_2\Bigr]
\left[ \ln(-p^2) + \ln(-x) \right] +  \frac{1}{2} \ln(x) \left[ \ln(-p^2)
\right.
\nonumber \\  &&
\left.
+ \ln(-x)
\right]^2
-\frac{\zeta_2}{2} \ln(x) +4 S_{1,2}(-x)-  2 {\litr}(-x)-
2\zeta_3 \Bigr) +\cdots  \Bigr\}
\nonumber
\end{eqnarray}

The vertex master integral $C_0(0,m,0,m^2,m^2,p^2)$ in $d$ dimensions
is finite for small $\eps$; it has been derived in \cite{Fleischer:2003rm}:
\ba
C_0(0,m,0,m^2,m^2,p^2)
&=& -e^{\epsilon \gamma_E}
(m^2)^{\frac{d}{2}-3}
\Gamma \left(2-\frac{d}{2}\right)~~
\Biggl\{\frac{1}
          {2(d-3)}
       \Fh21\Fpm{1,1}{\frac{d-1}{2}}
\nl&& 
-~ \frac{ \sqrt{\pi} ~ \Gamma \left(\frac{d-2}{2}\right)
         }
         {4 \Gamma \left(\frac{d-1}{2}\right)}
        \left(-\frac{p^2}{4m^2}\right)^{\frac{d-4}{2}}
~ \Fh21\Fpm{1,\frac{d-2}{2}}{\frac{d-1}{2}}\Biggr\}.
\nl
\ea
Concerning the expansion with respect to $\eps$, the two coefficients of the
$\Fh21$-functions depend on
$L_m$ and $\ln(-p^2)$, respectively.
By  eliminating $L_m$ according to
\ba
L_m=\ln(-p^2)-(2 \ln(1-x)-\ln(x))
\ea
we obtain
\begin{eqnarray}
C_0(0,m,0,m^2,m^2,p^2)
&=&
-\frac{1}{p^2~ b} \Bigl\{
 \frac{1}{2} \ln^2(x)  +2 \litwo(x) + 4 \zeta_2
-\eps~ \Bigl( \frac{1}{3} \ln(x)^3 -\frac{1}{2} \ln(x)^2
\nonumber \\&&
+[\ln(-p^2)-2\ln(1+x)] [
\frac{1}{2} \ln^2(x) +2 \litwo(x)+4\zeta_2 ]  \nonumber \\&&
 + \ln(x) [2 \litwo(x)-2\litwo( -x) +5 \zeta_2 ]  
- 2 S_{1,2}(x^2)+4 S_{1,2}(-x) \nonumber \\&& 
+8 S_{1,2}(x) +2 {\litr}(-x)- 5\zeta_3 \Bigr) \Bigr\}
+\ldots 
\end{eqnarray}
In terms of HPLs, the function reads for $m=1$ \cite{web-masters:2004nn}:
\ba
C_0(0,m,0,m^2,m^2,p^2)
&=&-~{\tt V3l1m[x]}
\nl&=&
\frac{x}{(1 - x^2)}~(4~\zeta_2 + H[0, 0, x] + 2~H[0, 1, x])
\nl&&
-
\frac{\eps~x}{(1 - x^2)}~\Bigl[5~\zeta_3 + 8~\zeta_2~H[-1, x] - \zeta_2~H[0, x]
+ 8~\zeta_2~H[1, x]
\nl&&
+
  2~H[-1, 0, 0, x] + 4~H[-1, 0, 1, x] + H[0, 0, 0, x]
\nl&&
+ 2~H[0, 0, 1, x] +
  2~H[0, 1, 0, x] + 4~H[0, 1, 1, x]
\nl&&
+ 2~H[1, 0, 0, x] +
  4~H[1, 0, 1, x]
\Bigl]
.
\ea

\subsection{\label{masterD} Four-point function}
In {\tt LoopTools} notation \cite{Hahn:1998yk}, the four-point master integral in $d$ dimensions with two photons
in the $s$-channel is :
\ba
\Bx(t,s) &=& D_0(m^2,m^2,m^2,m^2,t,s,m^2,0,m^2,0)
\nl
&=&
 \frac{e^{\epsilon \gamma_E}}{i\pi^{d/2}}
~\int\frac{d^dk}{k^2(k^2+2kp_4)(k+p_1+p_4)^2(k^2-2kp_3)} .
\ea
We first give the $\eps$-expansion obtained from a representation based on generalized hypergeometric functions; see Subsection \ref{boxhyper}.
Here we collect and complement results presented in \cite{Fleischer:2003rm} and \cite{Fleischer:2003bg}. Given the general result for the
box diagram in $d$ dimensions, the coefficients of the $\eps$-expansion are naturally
obtained in terms of one-dimensional integrals. Alternatively we consider in
Subsection \ref{boxhpl} the method of differential equations, which also yields the coefficients in terms of one-dimensional integrals.
These can, however,  systematically be presented in the form of generalized harmonic polylogarithms, which makes this form quite attractive if one prefers
`analytic' results.
Finally, in Subsection \ref{boxmb} we add a representation in terms of a two-fold Mellin-Barnes integral, which appears to be quite elegant and has the advantage that the integrand is free of singularities even in the physical domain.
\subsubsection{\label{boxhyper}Hypergeometric functions}
A closed expression for the box function valid in $d$ dimensions is
known from
\cite{Fleischer:2003rm}. In this case a first order difference equation
with respect to the dimension $d$ was solved.
\footnote{We just mention that in \cite{Fleischer:2003bg} also a
  Feynman parameter representation for
$D_0$ was given, including terms proportional to $\eps$.}
Other difference equations use as parameter the powers of the propagators
, see e.g. \cite{Laporta:2000dc,Laporta:2001dd,Laporta:2003jz}.
The general result of \cite{Fleischer:2003rm} reads:
\begin{eqnarray}
e^{-\epsilon \gamma_E}{\Bx}(t,s) =
 - \frac{4 m^{d-4}}{s(t-4m^2)} \Gamma \left(2-\frac{d}{2}\right)
 F_2\left(\frac{d-3}{2},1,1,\frac32, \frac{d-2}{2};
 \frac{t}{t-4m^2}, ~z \right)
 \nonumber \\
+\frac{4 m^{d-4}}{(d-3)s(t-4m^2)}\Gamma\left(2- \frac{d}{2}\right)
F^{1;2;1}_{1;1;0} \left[^{\frac{d-3}{2}:~ \frac{d-3}{2},~1;~~~~ 1;}
_{\frac{d-1}{2}:~~~~~~ \frac{d-2}{2};~~-;}~z,1-\frac{4m^2}{s}\right]
\nonumber \\
 -\frac{\sqrt{\pi} (-s)^{\frac{d-4}{2}}}{2^{d-4}m  \sqrt{s}}
~\frac{\Gamma\left(\frac{d-2}{2}\right) \Gamma\left(2-\frac{d}{2}\right)}
{(t-4m^2) \Gamma\left( \frac{d-1}{2}\right)}
F_1\left(\frac{d-3}{2},1,\frac12; \frac{d-1}{2};-\frac{u}{t-4m^2},1-\frac{s}{4m^2}
\right),
\non\\
\label{solution}
\end{eqnarray}
with
\begin{eqnarray}
z&=&\frac{4 m^2}{s} \frac{u}{t-4 m^2} > 0.
\end{eqnarray}
The two photons are in the $s$-channel.
Naturally the cuts of the diagram
are different for the $t$-channel case, which
means that the hypergeometric functions are to be evaluated in different
domains
of analyticity. In (\ref{solution}), e.g.,  the imaginary part of the
diagram
comes only from the coefficient $(-s)^{{(d-4)}/{2}}$ of $F_1$.

The Appell hypergeometric functions in terms of their integral
representations are:
\begin{eqnarray}
F_1\left(\frac{d-3}{2},1,\frac{1}{2},\frac{d-1}{2};x,y\right)=
 \frac{d-3}{2}\int_0^1\frac{dt~ t^{\frac{d-5}{2}}}{(1-t~ x)
\sqrt{1-t~ y}},
\label{F1}
\end{eqnarray}
\begin{eqnarray}
F_2\left(\frac{d-3}{2},1,1,\frac{3}{2},\frac{d-2}{2};x,z\right)=
  \int_0^1\frac{\frac{1}{2} dt} { \sqrt{1-t} (1-t x)^{\frac{d-3}{2}}}
{~}_2F_1(1,\frac{d-3}{2},\frac{d-2}{2},\frac{z}{1-t x}),
\non
\label{F2}
\\
\end{eqnarray}
and the Kamp\'e de F\'eriet function is:
\begin{eqnarray}
F^{1;2;1}_{1;1;0} \left[^{\frac{d-3}{2}:~ \frac{d-3}{2},~1;~~~~ 1;}
_{\frac{d-1}{2}:~~~~~~ \frac{d-2}{2};~~-;}~~z,y\right]
=\frac{d-3}{2}\int_0^1\frac{dt~ t^{\frac{d-5}{2}}}{1-t~ y}
{~}_2F_1(1,\frac{d-3}{2},\frac{d-2}{2},z~ t).
\label{KdF}
\end{eqnarray}
See \cite{mathworld} for (\ref{F1}),
\cite{Fleischer:2003bg} for (\ref{KdF}), and  (\ref{F2}) is obtained
from the double integral representation of the $F_2$-function \cite{Prudnikov:1986III}).
With these representations we can derive the neeeded $\eps$ expansion.
Due to $\Gamma(2-\frac{d}{2})$ in the prefactors of (\ref{solution}),
their  $\eps$-expansion has to be done up to order  ${\eps}^2$.
This can be performed by expanding the integrands.
The numerical evaluation of the one-dimensional integrals
of the $\eps$-terms works quite nicely in general. Nevertheless
partial analytic results can also be obtained, see e.g.
 (\ref{F1eps}) and (\ref{simple}) .
Based on \cite{Fleischer:2003bg} we also give an
expansion of the integrals for the limit of small masses,
i.e. $-t \gg 4 m^2$ (neglecting terms of $\mathcal{O}(m^2)$ and
$\mathcal{O}(m^2 \ln(m^2)$ ).

For the $F_1$-function the $\eps$-expansion is easy
except for the analytic integration
following the expansion. In \cite{Fleischer:2003rm} the analytic
integration has been performed for an $F_1$-function in which
one of the arguments is $\mathcal{O}(\eps)$ and in \cite{Fleischer:2003bg}
the corresponding transformation to obtain such a form
has been described in detail.
For the real part of $F_1$ we thus have:
\begin{eqnarray}
F_1\left(\frac{d-3}{2},1,\frac{1}{2},\frac{d-1}{2};x,y\right)
&=&
-\frac{m}{\sqrt{s} b}\frac{d-3}{2}
\Bigl[{\rm Re}
\Bigl\{
\ln(B)  \non \\
&&\left. \left.
-\eps \Bigl(\litwo(1-A B)+\litwo\left(1-\frac{B}{A}\right)
-2 \litwo(1-B)
\right. \right.\non \\ && \left. \left.
+\frac{1}{2} \ln^2 A +\pi^2 \Bigr)
\right. \right. \non \\
&&\left. \left.
+{\eps}^2 \left(
      {\rm Li}_3\left( \frac{A(1-AB)}{A-B}\right)
     -{\rm Li}_3\left( \frac{A(A-B)}{1-AB}\right)
\right. \right. \right. \non \\ &&
    +2{\rm Li}_3\left( \frac{A(1-B)}{1-AB}\right)
\left. \left. \left.
    -2{\rm Li}_3\left( \frac{A(1-B)}{A-B}\right)
\right. \right. \right. \non \\ &&\left. \left. \left.
    +2{\rm Li}_3\left( \frac{1-B}{A-B}\right)
    -2{\rm Li}_3\left( \frac{1-B}{1-AB}\right)
\right. \right. \right. \non \\
&&\left. \left. \left.
+ 2\Bigl[{\rm Li}_2\left( \frac{A(A-B)}{1-AB}\right)
          -{\rm Li}_2\left( \frac{A(1-B)}{1-AB}\right)
\right. \right. \right. \non \\
&&\left. \left.\left.
          +{\rm Li}_2\left( \frac{1-B}{A-B}\right)
          -{\rm Li}_2(-A)
    \Bigr] \ln(A)
\right. \right. \right. \non \\
&&\left. \left.
~~
+\left[\frac12\ln^2(A)-\zeta(2)\right]\ln\left(\frac{B-A}{1-AB}\right)
\right. \right. \non \\&& \left. \left.
-\frac{1}{6} \ln^3\left(\frac{B-A}{1-AB}\right)
\right. \right.  \non \\
&&~~\left.
+\frac12\ln (A)\ln^2 \left(\frac{B-A}{1-AB}\right)
 \right)  +O(\epsilon^3)
 \Bigr\} \Bigr],
\label{F1eps}
\end{eqnarray}
with
\begin{eqnarray}
A&=&x(s)=\frac{a-1}{a+1}\sim -\frac{m^2}{s}~< ~0,
\\
B&=&-x(t)=\frac{b-1}{b+1}\sim \frac{m^2}{t}~<0,
\\
a&=&\sqrt{\frac{s-4 m^2}{s}},
\\
 b&=&\sqrt{\frac{t}{t-4 m^2}}.
\end{eqnarray}
Abbreviating (\ref{F1eps}) as ($b \sim 1$)
\begin{eqnarray}
F_1\left(\frac{d-3}{2},1,\frac{1}{2},\frac{d-1}{2};x,y\right)=
-\frac{m}{\sqrt{s} }(d-3)
\left[F_1^0+\eps F_1^1 + {\eps}^2 F_1^2 \right],
\end{eqnarray}
we obtain from (\ref{F1eps}) in the limit of small masses
with $r={-t}/{s}, 0 \le r \le 1$:
\begin{eqnarray}
&&F_1^0=-\ln(\frac{-t}{m^2}),
\non \\
&&F_1^1=-\frac{1}{2} \ln^2(\frac{s}{m^2})-2\zeta(2)- \litwo(1-\frac{1}{r}),
\non\\
&&F_1^2=-\frac{1}{6} \ln^3(\frac{s}{m^2})-2 \zeta(2) \ln(\frac{s}{m^2})
-\litr(1-\frac{1}{r})-2 \zeta(3) .
\non \\
\end{eqnarray}
For the  $F_2$- and Kamp\'e de F\'eriet functions
the same hypergeometric function ${~}_2F_1$ needs to be
expanded:
\footnote{Again a {\tt FORM}
  code for the automatized derivation of the $\eps$-expansion
by  M. Kalmykov has been used.}
\begin{eqnarray}
{~}_2F_1(1,\frac{1}{2}-\eps,1-\eps,w)&=&
\frac{1}{\sqrt{1-w}}
\left\{1-2 \eps \ln(1+v) + 2 {\eps}^2 \left[\ln^2(1+v)+\litwo(-v)\right]\right\} \non \\
&\approx& \frac{1}{\sqrt{1-w}} \left\{ (1+v)^{-2 \eps}+2{\eps}^2\litwo(-v) \right\}+O( {\eps}^3 ),
\label{expand}
\end{eqnarray}
with
\begin{eqnarray}
v=\frac{1-\sqrt{1-w}}{1+\sqrt{1-w}}.
\end{eqnarray}
For the $F_2$-function we have to use
\begin{eqnarray}
w=\frac{z}{1-xt}=\frac{4 v}{ (1+v)^2 },
\end{eqnarray}
and correspondingly for the Kamp\'e de F\'eriet function
\begin{eqnarray}
w=zt=\frac{4 v}{ (1+v)^2 },
\end{eqnarray}
and further  in the integral (\ref{F2})
\begin{eqnarray}
\frac{(1+v)^{-2 \eps}} { (1-t~ x)^{-\eps}}
=(\frac{z}{4 v})^{\eps},
\end{eqnarray}
and in the integral (\ref{KdF})
\begin{eqnarray}
 (t)^{-\eps} (1+v)^{-2 \eps}
=(\frac {z}{4 v})^{\eps}.
\end{eqnarray}
It appears natural to introduce $v$ as integration variable.
But a more
precise numerical integration results from an elimination of the
singularity at $t=1$ in (\ref{F2}) by the transformation $1-t=u^2$.
We then have in the considered order for (\ref{F2}):
\begin{eqnarray}
F_2\left(\frac{d-3}{2},1,1,\frac{3}{2},\frac{d-2}{2};x,z\right)
\approx
    \int_{0}^{1}\frac{du~}
 { \sqrt{x u^2 + \frac{4 m^2}{s} } }
 [(\frac{z}{4 v})^{\eps}+2\eps^2 \litwo(-v)] ,
\label{F2v}
\end{eqnarray}
with
\begin{eqnarray}
\frac{z}{4 v}&=&\frac{x}{4}\left(
\sqrt{ u^2-\frac{4 m^2}{t}}+
\sqrt{ u^2+\frac{4 m^2}{s}(1-\frac{4 m^2}{t})}\right)^2~~<~~1 ,
\end{eqnarray}
and
\begin{eqnarray}
v~~&=&\frac{\sqrt{ u^2-\frac{4 m^2}{t}}-
\sqrt{ u^2+\frac{4 m^2}{s}(1-\frac{4 m^2}{t}) }}
{\sqrt{ u^2-\frac{4 m^2}{t}}+
\sqrt{ u^2+\frac{4 m^2}{s}(1-\frac{4 m^2}{t})}}~~>~~0.
\end{eqnarray}
For the following we write
\begin{eqnarray}
F_2\left(\frac{d-3}{2},1,1,\frac{3}{2},\frac{d-2}{2};x,z\right)=
F_2^0+\eps F_2^1+{\eps}^2 F_2^2 +\cdots ,
\label{F2formal}
\end{eqnarray}
where $F_2^0$ is obained as
\begin{eqnarray}
F_2^0=\frac{1}{ \sqrt{x}} \ln \left( \frac{1+\sqrt{\frac{x}{1-z}}}
{{1-\sqrt{\frac{x}{1-z}}}} \right),
\end{eqnarray}
and the higher orders must be calculated from (\ref{F2v}) numerically.
In the limit of small electron mass they are:
\begin{eqnarray}
&&F_2^0=  \ln( \frac{s}{m^2}) ,
\non \\
&&F_2^1=-\frac{1}{2} \ln^2( \frac{s}{m^2})+
\zeta(2)-\litwo(1-\frac{1}{r}) ,
 \non\\
&&F_2^2= \frac{1}{6} \ln^3(\frac{s}{m^2})-\ln(\frac{s}{m^2})
\left(\zeta(2)-\litwo(1-\frac{1}{r})\right)+\frac{1}{2}\zeta(3)+
S_{1,2}(1-\frac{1}{r})-\litr(1-\frac{1}{r}) .
\nonumber \\
\end{eqnarray}
Similarly we perform the calculation for the Kamp\'e de F\'eriet function (\ref{KdF}):
\begin{eqnarray}
&&F^{1;2;1}_{1;1;0} \left[^{\frac{d-3}{2}:~ \frac{d-3}{2},~1;~~~~ 1;}
_{\frac{d-1}{2}:~~~~~~ \frac{d-2}{2};~~-;}~~z,y\right] \sim
-\frac{d-3}{2}\frac{1}{ \sqrt{y-z}} \non \\
&&\int_0^1
du \left\{\left[\frac{1}{1+b_1 u} + \frac{1}{1-b_1 u} \right] b_1 -
\left[\frac{1}{1+b_2 u} + \frac{1}{1-b_2 u} \right] b_2\right\}
\left[ (\frac{z}{4 v})^{\eps} +2\eps^2 \litwo(-v)\right]  \non \\
\label{KdFz}
\end{eqnarray}
with $v=v_0 u^2$ and
\begin{eqnarray}
v_0&=&\frac{1-\sqrt{1-z}}{1+\sqrt{1-z}} \sim \frac{z}{4}, ~~~
v_1=\frac{1+\sqrt{1-\frac{z}{y} } }{1-\sqrt{1-\frac{z}{y} } }, ~~~ 
v_2=\frac{1}{v_1}, \non \\
b_1&=&\sqrt{\frac{v_0}{v_1}} \ll 1~~~~~{\rm and} ~~~~~~
b_2=\sqrt{\frac{v_0}{v_2}}=\sqrt{v_0 v_1}.
\end{eqnarray}
As above for the $F_2$, we formally write for the Kamp\'e de F\'eriet
function:
\begin{eqnarray}
F^{1;2;1}_{1;1;0} \left[^{\frac{d-3}{2}:~ \frac{d-3}{2},~1;~~~~ 1;}
_{\frac{d-1}{2}:~~~~~~ \frac{d-2}{2};~~-;}~~z,y\right] =
\frac{d-3}{2}\left( K^0+\eps K^1+{\eps}^2 K^2 +\cdots \right) ,
\label{KdFformal}
\end{eqnarray}
where $K^0$ is obtained as
\begin{eqnarray}
K^0=\frac{1}{b} \ln \left( \frac{(1-b_1) (1+b_2)}{(1+b_1) (1-b_2)} \right)  .
\end{eqnarray}
Again, investigating the small mass approximation, we have
\begin{eqnarray}
K^0=&&\ln ( \frac{s}{m^2} ) ,
\non \\
K^1=&&3 \zeta(2),
\non \\
K^2=&&7 \zeta(3).
\end{eqnarray}
Finally we see that the expansion in $\eps$ of the
$F_2$- and Kamp\'e de F\'eriet functions becomes easy
with the representations (\ref{F2v}) and (\ref{KdFz}).
To sum up our results, we have
\begin{eqnarray}
\Bx(t,s) =
&& - \frac{2 (m^2)^{-\eps}}{s(t-4m^2)} \Gamma \left(\eps \right)
\left[F_2^0+\eps F_2^1+{\eps}^2 F_2^2 +\cdots \right]
 \nonumber \\
&&+\frac{2 (m^2)^{-\eps} }{s(t-4m^2)}\Gamma\left(\eps \right)
\left[ K^0+\eps K^1+{\eps}^2 K^2 +\cdots  \right]
\nonumber \\
&&+\frac{2(\frac{-s}{4})^{-\eps} }{s(t-4m^2)}\Gamma\left(\eps \right) \frac{\Gamma\left(\frac{1}{2}\right)}{\Gamma\left(\frac{1}{2}-\eps\right)}\Gamma \left(1-\eps \right)
\left[{\rm Re} \left\{F_1^0+\eps F_1^1+{\eps}^2 F_1^2 +\cdots \right\} \right] .
\non
\\
\label{expso}
\end{eqnarray}
As we see, in the limit of small electron mass the ${1}/{\eps}$-terms of the $F_2$- and Kamp\'e de F\'eriet functions cancel.

It is a very appealing fact to have the closed form of the box function as an analytical expression in $d$ dimensions.
So far, however,  only partial analytic results were obtained for the
terms of order $\eps$ , but as we observe already from ({\ref{F1eps}}),
the results become quite lengthy if one prefers to present them
in this form. Beyond that simple expressions in the small mass
limit were obtained.
\subsubsection{\label{boxhpl}Harmonic polylogarithms}
 An alternative approach
in terms of solving a differential equation for the box \cite{Bonciani:2003cj} yields in a natural
manner harmonic polylogarithms.
For the purpose of checking (in particular also numerically) and comparing, we repeated the calculation
of \cite{Bonciani:2003cj} and shortly sketch the procedure.

To be explicit, we consider the Bhabha box diagram with two photons in the $t$-channel,
as in \cite{Bonciani:2003cj}, and the electron mass being set to 1;
the analytical continuation to the $s$-channel is evident here.
One may derive the differential operator
\bea
s \frac{\partial}{\partial s}&=& \frac{1}{2}
\left\{ p_1^{\mu}+p_2^{\mu}+\frac{s}{s+t-4} (p_2^{\mu}-p_3^{\mu})
\right\} \frac{\partial}{\partial p_2^{\mu}}~~~,
\eea
which applied to the one-loop box yields a
differential equation:
\bea
\frac{\rd \Bx(s,t)}{\rd s} &=& \frac{1}{2(-4+s)^2 s t (-4+s+t)}
\non\\
&& \Biggl[ (-4+s)t(-2 s^2+4(-4+t)+s(12+(-6+\rd)t))~\Bx(x,y)
\non\\
&&-  2(-4+s)(-4+\rd)(-4+t)t ~{\VVtt} (y)
\non\\
&&
+4st(-3+\rd)~{\tt SE2l2m}(x)
\non\\
&& -4(-3+\rd)(-4+s)(-4+s+t)~{\tt SE2l0m}(y)
\non\\
&&
-2(-2+\rd)st~{\tt T1l1m}\Biggr],
\label{Diff}
\eea
where
\bea
x&=&\frac{\sqrt{1-4/s}-1}{\sqrt{1-4/s}+1},
\\
 y&=&\frac{\sqrt{1-4/t}-1}{\sqrt{1-4/t}+1},
\eea
or
\bea
s&=&-\frac{(1-x)^2}{x},
\\
t&=&-\frac{(1-y)^2}{y}.
\eea
The subdiagrams {\tt T1l1m}, {\tt SE2l2m}, {\tt SE2l0m}, {\tt V3l1m} are given in the preceding sections.

Expanding now the differential equation (\ref{Diff}) in $\eps$ and introducing the ansatz
\bea
\Bx
&=&\mathrm{const}~ {\tt B4l2m}(x,y)
\nl
&=&\frac{1}{\eps} B_{-1}+B_0+\eps B_1 + \cdots,
\eea
we may iteratively solve a system of differential equations which
differ only in the inhomogeneous terms:
\bea
\frac{\rd B_j(x,y)}{\rd x} =\frac{1+x^2}{x(1-x^2)} B_j(x,y)+ C_j(x,y).
\label{Diffj}
\eea
More details are described in the literature, e.g. in \cite{Bonciani:2003cj}.

The result is
\be
B_{-1}=\frac{2 x y H(0,x)}{(1-x^2)(1-y)^2} =\frac{2}{ s t\sqrt{1-4/s}}
H(0,x),
\label{Bminus}
\ee
where $H(0,x) \equiv \ln(x)$ has been introduced,
\be
B_{0}=
\frac{2}{ s t\sqrt{1-4/s}}  H(0,x)\Bigl( H(0,y)+2 H(1,y) \Bigr) ,
\label{Bnull}
\ee
and
finally
\bea
B_1
&=& \frac{-2}{ s t\sqrt{1-4/s}} \left\{G(-\frac{1}{y},0,0,x)+G(-y,0,0,x) \right. \non \\
&& \left. -2 \Bigl(G(-\frac{1}{y},-1,0,x)+G(-y,-1,0,x)\Bigr)  \right. \non \\
&& \left. - \Bigl(G(-\frac{1}{y},0,x)+G(-y,0,x)-2 H(-1,0,x)\Bigr)[H(0,y)+2 H(1,y)] \right. \non \\
&& \left. - \Bigl(G(-\frac{1}{y},x)-G(-y,x)+ H(0,x)\Bigr)[H(0,0,y)+2 H(0,1,y)] \right.  \non \\
&& \left. - \Bigl(5 G(-\frac{1}{y},x) -3 G(-y,x)-\frac{1}{2} H(0,x)-2  H(-1,x) -4 H(0,y)\Bigr)
             \zeta_2 \right. \non \\
&& \left. -2 \Bigl(H(1,y) H(0,0,y)-H(0,y) H(0,1,y)\Bigr)  \right. \non \\
&& \left. -2\Bigl( H(-1,0,0,x)-2 H(-1,-1,0,x)\Bigr) -2 H(0,x)[H(1,0,y)+2 H(1,1,y) ] \right. \non \\
&& \left. +H(0,0,0,y)+2 H(1,0,0,y)  -2 \zeta_3 ~~ \right\} .
\label{B1}
\eea
The functions $G$ are generalized harmonic polylogarithms \cite{Gehrmann:2000zt,Bonciani:2003cj}.
For the calculation of $B_1$ we used the relations
\bea
 G(-y,0,0,1) &=&  -\zeta_2H(0,y)+H(0,0,-1,y)-H(0,0,0,y),
  \\
 G(-y,-1,0,1)+G(-\frac{1}{y},-1,0,1) &=&
 -\frac{3}{2}\zeta_3-\zeta_2H(-1,y)+ H(1,0,0,y)+H(0,0,-1,y).
\non\\
\eea
There is a difference in the coefficient of the term $H(0,x) \zeta_2$
w.r.t. \cite{Bonciani:2003cj} due to different choices of normalization, see also
(\ref{Bminus}).

In order to check the results, we evaluated both representations of $B_1$ numerically
(for the photons in the $s$-channel).
For $s=10^6, \cos\theta = 0.4, m = 1$, agreement to nine decimals was achieved:
\bea
B_1(s=10^6,\cos\theta=0.4)=4.43779985~ 10^{-9}-1.61529999~ 10^{-9}~i.
\label{num}
\eea
The most difficult part of the numerical evaluation of (\ref{B1}) is the calculation of $G(-y,-1,0,x)$,
in which case a principal value integral has to be performed with the above parameters. The
imaginary part obtained from (\ref{B1}) therefore agrees only to 7 decimals with (\ref{num}).
Following \cite{Fleischer:2003bg}, a simple formula for the imaginary part of $B_1$ can be derived, which is indeed simpler than what is obtained from (\ref{B1}):
\bea
\Im(B_1)&=&\frac{\pi}{ s \sqrt{t(t-4)}} \non
\\
&&    \Bigl( 2 \litwo(1+xy) +2 \litwo(1+y/x) + 4\litwo(-y) +\ln^2(-x) +\frac{{\pi}^2}{3}
 \non \\
&&  +2 \ln(y)[\ln(s)+2\ln(1+y)] \Bigr).
\label{simple}
\eea
This yields the imaginary part of the above number.
The numerical calculations were performed with {\tt Mathematica} and
{\tt Maple}, respectively.

\subsubsection{\label{boxmb}Mellin-Barnes representation}
\def \Log {{\rm{ln}}}
Finally, we derive a Mellin-Barnes representation for the QED box integral, again with two photons in the $s$-channel.
The  Mellin-Barnes representation reads for finite $\eps$:
\begin{eqnarray}
{\tt Box}(t,s) &=&
   \frac{e^{\epsilon \gamma_E}}{ \Gamma[ - 2\eps ] (-t)^{(2+\eps)}}\frac{1}{(2 \pi i )^2}
 \int_{-i \infty}^{+i \infty} d {z_1} \int_{-i \infty}^{+i \infty}
d {z_2}
\\\nonumber&&
   \frac{(-s)^{z_1}(m^2)^{z_2}}{(-t)^{z_1+z_2}}
\Gamma[2  + \eps+ z_1+z_2] {\Gamma}^2[1+z_1]
 \Gamma[-z_1]\Gamma[-z_2]
   \\\nonumber&&
\Gamma^2[-1  - \eps -z_1  - z_2 ]
  \frac{\Gamma[-2 - 2\eps  - 2 z_1]}
{ \Gamma[-2 - 2\eps- 2 z_1  -2 z_2 ]}
\end{eqnarray}
A derivation may be found e.g. in \cite{Smirnov:2004}.
Starting from this Mellin-Barnes integral, one has to perform an
analytic
continuation in $\epsilon$ from a domain where the integral is regular
into the vicinity of the origin. The singularity structure near
$\eps \sim 0$ is obtained by means of the {\tt Mathematica} package
{\tt MB} \cite{Czakon:2005rk}.
We obtain the result in terms of the following one- and two-
dimensional integrals:
\begin{eqnarray}
{\tt I1}= \frac{e^{\epsilon \gamma_E}}{s t} \frac{1}{2 \pi i }
    \int_{-\frac{1}{2} -i \infty}^{-\frac{1}{2}+i \infty} d {z_1}
  \left(\frac{m^2}{-t}\right)^{z_1}
\frac{{\Gamma}^3[-z_1] \Gamma[1+z_1]}
 { \Gamma[- 2 z_1]},
\label{Int1}
\end{eqnarray}
and
\begin{eqnarray}
{\tt I2}= \frac{e^{\epsilon \gamma_E}}{t^2}\frac{1}{(2 \pi i )^2}
    &&\int_{-\frac{3}{4} -i \infty}^{-\frac{3}{4}+i \infty} d {z_1}
 \left(\frac{-s}{-t}\right)^{z_1}\Gamma[-z_1]\Gamma[-2(1+z_1)]
{\Gamma}^2[1+z_1]
\\\nonumber
    &&\times \int_{-\frac{1}{2} -i \infty}^{-\frac{1}{2}+i \infty} d {z_2}
\left(\frac{m^2}{-t}\right)^{z_2}\Gamma[-z_2]
\frac{{\Gamma}^2[-1-z_1-z_2]}{\Gamma[-2(1+z_1+z_2)]}
  { \Gamma[ 2 + z_1 + z_2]}.
\label{Int2}
\end{eqnarray}
In terms of the conformally mapped variable
\be
y=\frac{\sqrt{1-4 m^2/t}-1}{\sqrt{1-4 m^2/t}+1},
\ee
the first integral ${\tt I1}$ in (\ref{Int1}) can be performed analytically to yield the well
known result
\begin{eqnarray}
{\tt I1}= \frac{1}{ m^2 s} \frac{2 y }{1 - y^2} \Log(y).
\label{Inta}
\end{eqnarray}
The final result for the ${\tt Box}$ then reads:
\begin{eqnarray}
{\tt Box}(t,s) =-\frac{1}{\eps}{\tt I1}+\Log(-s){\tt I1}+
\eps \left(\frac{1}{2}\left[\zeta(2)-{\Log}^2(-s)\right]{\tt I1}
-2 {\tt I2}\right) .
\end{eqnarray}
The first two terms are in evident agreement with (\ref{Bminus}) and (\ref{Bnull}).
The double integral {\tt I2} in (\ref{Int2}) is not easily evaluated analytically,
although we know the answer from (\ref{B1}).
The MB package yields fairly precise
values in the Euclidean region ($s < 0$). In the Minkowskian
domain (with $s > 0$ and $(-s)^{z_1}=s^{z_1} exp(-i \pi z_1)$)
our experience with {\tt Mathematica} is that the built-in function
{\tt NIntegrate} with {\tt MaxRecursion $\rightarrow$ 12} gives easily a precision of nine decimals.
An alternative is the expansion at small  $m$ and fixed value of $t$.
With
\begin{eqnarray}
m_t&=&\frac{-m^2}{t},
\\
r&=&\frac{s}{t} ,
\end{eqnarray}
we have obtained a compact answer for {\tt I2} with the additional aid of {\tt XSUMMER} \cite{Moch:2005uc}.
The box contribution in this limit  becomes:
\begin{eqnarray}
B_1 &=&  \frac{1}{st} \Bigl\{ 4 \zeta_3 -9 \zeta_2 \Log(m_t)+
\frac{2}{3} {\Log}^3(m_t) + 6 \zeta_2 \Log(r)-{\Log}^2(m_t)\Log(r)
\\ \nonumber
&&+  \frac{1}{3} {\Log}^3(r) - 6 \zeta_2 \Log(1+r) +
2 \Log(-r)\Log(r)\Log(1+r)-{\Log}^2(r) \Log(1+r)
\\ \nonumber
&&
+ 2 \Log(r) \litwo(1+r)+2 \litr(-r) \Bigr\} +{\mathcal O}(m_t).
\end{eqnarray}

\section{\label{summary}Summary}
A calculation of
Bhabha scattering for the luminosity measurement at ILC is promoted by several groups, aiming at
a precision of $0.01\%$.
With this study, we provide a publicly available program for the
one-loop electroweak Standard Model corrections.
Further we collect all needed expressions for the factorizing one-loop QED corrections.
They are necessary ingredients for the full
two-loop calculation of Bhabha scattering.

\section*{Acknowledgements}
\noindent
We would like to thank O. Tarasov and A. Werthenbach for cooperation at an early stage of the project and
M. Kalmykov for discussions, some checks, and his {\tt FORM} code for the $\eps$-expansion of hypergeometric functions prior to publication.
J.F. is grateful to DESY, Zeuthen for support.
\def\thesection{\Alph{section}}
\def\theequation{\thesection.\arabic{equation}}
\setcounter{section}{0}
\setcounter{equation}{0}
\section{ \label{reduction}Reduction of tensor and scalar loop functions to master integrals}
We strictly apply here dimensional regularization, i.e.
infrared as well as ultraviolet singularities are given in terms of only one
pole in $\epsilon = (4 - d )/2 $.

After using {\tt DIANA} \cite{Tentyukov:1999is} and {\tt FORM} 3.1 in order to
express the Feynman diagrams in terms of tensor integrals,
we have to express the latter ones by scalar integrals: writing
the $m^{th}$ scalar amplitude of a diagram
formally as
\ba
F_m^{Diagram} &=&  \sum_{n,l} I_{n,l}^0 +  \sum_{i,n,l} p_{i,\mu} I_{n,l}^{\mu} +  \sum_{i,j,n,l} p_{i,\mu} p_{j,\nu} I_{n,l}^{\mu\nu}
\label{dots}
\ea
where momenta $p_{i,\mu}$ are the `chords', i.e. momenta in the propagators  $ c_i = (k-p_i)^2-m_i^2$, with $k$ the loop momentum.
The generically denoted $n$-point scalar, vector, and
tensor integrals $I_{n,l}^0, I_{n,l}^{\mu}, I_{n,l}^{\mu, \nu}$
will be
transformed into scalar integrals
with shifted space-time dimension $d$, which are then reduced to scalar integrals in generic dimension
by means of recursion relations \cite{Davydychev:1991va,Tarasov:1996br,Fleischer:1999hq}.
The indices $l$ in (\ref{dots}) stand
for  `dots' on lines $l$.
The reduction to scalar integrals reads:
\begin{eqnarray}
  I_{n,j}^{\mu} & =& \int ^{d} k_{\mu} \prod_{r=1}^{n} \, {c_r^{-(1+{\delta}_{rj})}}  =
   -\sum_{i=1}^{n-1} \, p_i^{\mu} \, n_{ij} \, I_{n,ij}^{[d+]} ,
   \nn\\
 I_{n,l}^{\mu\, \nu}& =& \int ^{d} k_{\mu} \, k_{\nu} \, \prod_{r=1}^{n} \, {c_r^{-(1+{\delta}_{rl})}}
  =  \sum_{i,j=1}^{n-1} \, p_i^{\mu}\, p_j^{\nu} \, n_{ijl} \,  \, I_{n,ijl}^{[d+]^2} -\frac{1}{2}
   \, g^{\mu \nu}  \, I_{n,l}^{[d+]} \, ,
\label{intdot}
\end{eqnarray}
where  $[d+]$ is an operator shifting the space-time dimension by two units ,~ $n_{ij}=(1+{\delta}_{ij})!, ~ n_{ijl}=(1+{\delta}_{ij}+{\delta}_{il}
+{\delta}_{jl}-{\delta}_{ij} {\delta}_{il} {\delta}_{jl})!$ and
\begin{eqnarray}
  \label{eq:Inij}
   I_{n, \, i\,j...} =  \int ^{d}  \prod_{r=1}^{n} \, \frac{1}{c_r^{1+\delta_{ri} + \delta_{rj}...}},~~~ \int ^{d} \equiv \int \frac{d^d k}{\pi^{d/2}}
\end{eqnarray}
is the original scalar integral with additional powers (dots) of the $i$-th and  $j$-th propagators. The case with no dots is formally obtained
by putting $j=l=0$.
 Having reduced the tensor integrals to scalar integrals, the generic space-time dimension $ d $  needs to be re-established and the dots
to be removed.
For this we use the recurrence relations first proposed in \cite{Tarasov:1996br}, which are complementary to those obtained via integration by parts \cite{Tkachov:1981wb,Chetyrkin:1981qh}, and later
  simplified and extended to zero Gram determinants in
\cite{Fleischer:1999hq}.
With $ Y_{ij}=-(p_i-p_j)^2+m_i^2+m_j^2$ and
the Cayley determinant
\begin{eqnarray}
()_n ~\equiv~  \left|
\begin{array}{ccccc}
  0 & 1       & 1       &\ldots & 1      \\
  1 & Y_{11}  & Y_{12}  &\ldots & Y_{1n} \\
  1 & Y_{12}  & Y_{22}  &\ldots & Y_{2n} \\
  \vdots  & \vdots  & \vdots  &\ddots & \vdots \\
  1 & Y_{1n}  & Y_{2n}  &\ldots & Y_{nn}
\end{array}
\right|,
\end{eqnarray}
the so-called signed minors  $ {j_1\, j_2 ... \choose i_1 \, i_2 ...}_n $ are determinants
where the
 rows $j_1, j_2, ...$ and columns $ i_1, i_2, ...$ are erased from the Cayley determinant $()_n$.\footnote{Note here the additional overall sign  $(-1) ^{j_1+j_2+...+i_1+i_2+..} $. } Making successive use of the following three
recurrence relations leads to scalar master integrals  $A_0, B_0, C_0 $ and  $D_0 $ in  $d $ dimensions:
{\small
\begin{eqnarray}
\left(  \right)_n
 \nu_j{\bf j^+} I^{(d+2)}_n  &&=
\left[  - {j \choose 0}_n +\sum_{k=1}^{n} {j \choose k}_n
 {\bf k^-} \right] I^{(d)}_n ,
\label{recur1}
\\
{0\choose 0}_n \nu_j {\bf j^+} I_n^{(d)} &&=
   \left[ \left( 1 + \sum_{i=1}^n \nu_i - d \right) {0\choose j}_n \! -
\sum_{k=1}^n  {0j\choose 0k}_n \! (\nu_k-1)   \right] \, I_n^{(d)}
\non \\[2mm]
&&- \sum_{i,k \, i\neq k}^n   {0j\choose 0k}_n \,\, \nu_i \, {\bf
k^-} {\bf i^+}\, I_n^{(d)}  ,
\label{recur2}
\\
  (d-\sum_{i=1}^{n}\nu_i+1) \left(  \right)_n  I^{(d+2)}_n && =
    \left[ {0 \choose 0}_n - \sum_{k=1}^n {0 \choose k}_n {\bf k^-} \right]I^{(d)}_n \, .
\label{recur3}
\end{eqnarray}}
These relations are applied in a {\tt FORM} program one after the other: the (\ref{recur1})
reduces the dimension and the index of the $j^{th}$ line, the
(\ref{recur2}) reduces
the index of the $j^{th}$ line without changing the space-time dimension.
The third relation (\ref{recur3}) also reduces the space-time.
The operators ${\bf i^+, j^+} $ raise the power of the corresponding propagator by one unit, while ${\bf k^-} $ reduces the power of the  $k $-th propagator by one unit.

For Bhabha scattering in particular there is one subtlety: there occur zero
Gram determinants and for this case special care must be taken. \noindent
The occurrence of zero Gram determinants  (e.g. $()_n=0 $) is discussed  in \cite{Fleischer:1999hq}.
Effectively a zero Gram determinant reflects the kinematical boundaries of phase space where a given
$n$-point function can be
expressed through scalar integrals of lower rank. A typical example of such simplifications is
\begin{eqnarray}
C_0(m,0,m;m^2,m^2,s) &=&
 \frac{1}{s-4m^2}
\left[\frac{d-2}{d-4}~\frac{A_0(m^2)}{m^2}
     + \frac{2d-3}{d-4}~ B_0(m,m;s)\right].
\nn\\
\label{C4}
\end{eqnarray}
(\ref{C4}) means that this $C_0$ is
strictly speaking not a master integral. For practical reasons, however,
we include it in the list of master integrals: see the discussion
in Section \ref{masterC}.
It is
instructive to derive (\ref{C4})
in order to demonstrate how the procedure works.
Setting the momenta of the incoming massive lines to $p_1$ and $p_2$
(the third momentum $q=-p_1-p_2$, $q^2=s$) and the integration momentum
on the massless line (no. 3), then the chords are, respectively, $-p_1, p_2$ and
$0$. Correspondingly we have for the Cayley determinant
\begin{eqnarray}
()_3 ~\equiv~  \left|
\begin{array}{ccccc}
  0 & 1       & 1         & 1 \\
  1 &    2 m^2& -s+2 m^2  & 0 \\
  1 & -s+2 m^2&    2 m^2  & 0 \\
  1 &     0   &     0     & 0
\end{array}
\right| =s(s-4 m^2),
\end{eqnarray}
Apparently ${0\choose 0}_3=0$.
 Applying (\ref{recur3}) with $d \to d-2$,
we obtain
\begin{eqnarray}
(d-4) \left(  \right)_3  I^{d}_3 =
- \sum_{k=1}^3 {0 \choose k}_3 {\bf k^-} I^{(d-2)}_3 ,
\label{recur0}
\end{eqnarray}
where ${0\choose 1}_3={0\choose 2}_3=0$ and ${0\choose 3}_3=-s(s-4 m^2)$.

Now we have expressed our three point function already in terms of
a two point function with two massive lines, however in $d-2$
dimensions, i.e.
\begin{eqnarray}
(d-4) I^{d}_3 = {\bf 3^-} I^{(d-2)}_3
\end{eqnarray}
 and we must increase the dimension again with the intention
to obtain an integral with nonva- nishing Gram determinant. The relevant
relation to be used is (29) in \cite{Fleischer:1999hq}
\begin{eqnarray}
\sum_{j=1}^n {\nu}_j {\bf j^+} I^{(d+2)}_n=-I^{d}_n,
\end{eqnarray}
which in our case yields
\begin{eqnarray}
I^{(d-2)}_2=-\sum_{j=1}^2 {\bf j^+} I^{d}_2=-2~~ {\bf 1^+} I^{d}_2  ,
\end{eqnarray}
i.e. a two point function in generic dimension with a dot on one
of the two massive lines and we have to remove the dot from the line.
In this case we have $()_2=- 2 s$ and ${0\choose 0}_2=-s(s-4 m^2)$,
i.e. both Gram determinants are nonvanishing and we can apply
(\ref{recur2}), which yields straightforwardly (\ref{C4}).


\providecommand{\href}[2]{#2}


\end{document}

%% file: macro.tex
\def\sigmap{\sigma^{\prime}}
\def\mup{\mu^{\prime}}
\def\nup{\nu^{\prime}}
\def\rhop{\rho^{\, \prime}}
\def\bb{b \bar{b}} 
\def\cc{c \bar{c}} 
\def\qq{q \bar{q}} 
\def\cM{{\cal M}} 
\def\cO{{\cal O}}
\def\cK{{\cal K}} 
\def \ni {\noindent}
\def \be {\begin{equation}}
\def \e {\end{equation}}
\def \bea {\begin{eqnarray}}
\def \ea {\end{eqnarray}}
\def \eps {\epsilon}
\def \si {\sigma}
\def \ga {\gamma}
\def \ka {\kappa}
\def \la {\lambda}
\def \no {\nonumber}
\def \G {{\rm g}}
\def \dd {{\rm d}}
\def \Li {{\rm Li_2}}
\def \K {k_{\bot}^2}
\def \Vec#1{\mbox {\bf #1}}
\def \Veg#1{\mbox{\boldmath $#1$}}
\def \sub {\scriptscriptstyle}
\def \ps {p\hspace{-0.43em}/}
\def \mps {p\hspace{-0.45em}/}
\def \sps {p\hspace{-0.32em}/}
\def \ns {n\hspace{-0.51em}/}
\def \mns {n\hspace{-0.53em}/}
\def \ks {k\hspace{-0.49em}/}
\def \es {\epsilon\hspace{-0.47em}/}
\def \Z#1#2{\widetilde{Z}^{#1}_{#2}}
\def \Zw{\ensuremath{ {Z^{-\frac{1}{2}}_{\scriptscriptstyle W}\, } }}
\def \Zz{\ensuremath{ {Z^{-\frac{1}{2}}_{\scriptscriptstyle Z}\, } }}
\def \Zwn{\ensuremath{ \delta Z_{\scriptscriptstyle W,\,n} \,}}
\def \Zzn{\ensuremath{ \delta Z_{\scriptscriptstyle Z,\,n} \,}}
\def \slash#1{#1 \hspace{-0.42em}/}
\newcommand{\To}[2]{\stackrel{#1}{\hbox to #2 pt{\rightarrowfill}}}

\def \an {\widehat}
\def \abs#1{|\,#1\,|}
\def \vector#1{\stackrel{\hspace{-0.45em}\longrightarrow}{#1}}
\def \pa {\partial}
\def \c {\hspace{-0.2em} \cdot}
\def\wp{\ifmmode W^+\else $W^+$\fi}
\def\wm{\ifmmode W^-\else $W^-$\fi}
\def\emm{\ifmmode e^-\else $e^-$\fi}
\def\ep{\ifmmode e^+\else $e^+$\fi}

\def\sw{\ensuremath{ \sin \theta_{\rm w}}}
\def\swto{\ensuremath{ \sin^2 \theta_{\rm w}}} 
\def\swfor{\ensuremath{ \sin^4 \theta_{\rm w}}} 
\def\cw{\ensuremath{ \cos \theta_{\rm w}}} 
\def\cwto{\ensuremath{ \cos^2 \theta_{\rm w}}}
\def\cwfor{\ensuremath{ \cos^4 \theta_{\rm w}}}

\def\ie{{\it i.e.~}}
\def\eg{{\it e.g.~}}

\def\tw{\ensuremath{ \tan \theta_{\rm w} }}
\def\ctw{\ensuremath{ \cot \theta_{\rm w} }}
\def\twto{\ensuremath{ \tan^2 \theta_{\rm w} }}
\def\ctwto{\ensuremath{ \cot^2 \theta_{\rm w} }}

\def\mw{\ensuremath{ {M}_{\scriptscriptstyle W}} }
\def\mz{\ensuremath{ {M}_{\scriptscriptstyle Z}} }
\def\mh{\ensuremath{ {M}_{\scriptscriptstyle H}} }
\def\mn{\ensuremath{ {M}_{\scriptscriptstyle N}} }
\def\mwp{\ensuremath{ {M}_{{\scriptscriptstyle W},\,{\rm phys. }}}}
\def\mzp{\ensuremath{ {M}_{{\scriptscriptstyle Z},\,{\rm phys. }}}}

\def \mt{\ensuremath{ m_t}}

\def\B#1{\ensuremath{\delta Z^{\, (1)}_{W_T}({#1})   }}
\def\S#1{\ensuremath{\delta Z^{\, (1)}_{\phi}({#1})   }}

\def \Cw#1{\ensuremath { \delta Z^{\, (1)}_{f_{#1}}(W)}}
\def \Cz#1{\ensuremath { \delta Z^{\, (1)}_{f_{#1}}(Z)}}
\def \Cga#1{\ensuremath { \delta Z^{\, (1)}_{f_{#1}}(\ga) }}

\def\L{\ensuremath{ {\rm L}}}
\def\z#1{\ensuremath { \frac{\dd z_{#1}}{z_{#1}}  }}
\def\y#1{\ensuremath { \frac{\dd y_{#1}}{y_{#1}}  }}

\def \onshellm{\vphantom{\frac{1}{1}}_{\left|_{\scr{\,p^2=m^2}}\right.}}
\def \onshell{\vphantom{\frac{1}{1}}_{\left|_{\scr{\,k^2=\mw^2}}\right.}}
\def \onshellW{\vphantom{\frac{A^2}{A^2}}\left|_{_{\scr{\,k^2=\mw^2}}}\right.}
\def \onshellZ{\vphantom{\frac{A^2}{A^2}}\left|_{_{\scr{\,p^2=\mz^2}}}\right.}
\def \onshellA{\vphantom{\frac{A^2}{A^2}}\left|_{_{\scr{\,p^2=0}}}\right.}
\def \onshellAIR{\vphantom{\frac{A^2}{A^2}}\left|_{_{\scr{\,k^2 \to 0}}}\right.}

\def \onshellH{\vphantom{\frac{A^2}{A^2}}\left|_{_{\scr{\,k^2=\mh^2}}}\right.}
\def \onshellf{\vphantom{\frac{A^2}{A^2}}\left|_{_{\scr{\,p^2=m_f^2}}}\right.}
\def \onshellp{\vphantom{\frac{1}{1}}_{\left|_{\scr{\,k^2=\mwp^2}}\right.}}
\def \onshellWp{\vphantom{\frac{A^2}{A^2}}\left|_{_{\scr{\,k^2=\mwp^2}}}\right.}

\def \dd#1{\frac{\partial}{\partial \, {#1}^2}}

\def \h#1{ \hspace*{#1mm}}
\def \v#1{ \vspace*{#1mm}}

\def\dis#1{\ensuremath { {\displaystyle  #1}}}  
\def\scr#1{\ensuremath { { \scriptstyle #1}}}  
\def\sscr#1{\ensuremath { { \scriptscriptstyle #1}}}  

\def \myto#1#2{\ensuremath {\begin{array}[H]{c}
 {\scriptstyle {\rm #1}} \\[-2mm] {-\!\!\!-\!\!\!-\!\!\!-\!\!\!\longrightarrow} \\[-2mm] {\scriptstyle {\rm #2}}
\end{array} }}

\def \myk{\ensuremath {\begin{array}[H]{c} \\[-7mm] {-\!\!\!-\!\!\!\longrightarrow} \\[-2mm] k
\end{array} }}

\def \Myto#1{\ensuremath {\begin{array}[H]{c}
 {\scriptstyle { #1}} \\[-2mm] {-\!\!\!\longrightarrow} \end{array} }}
\def \Myeq#1{\ensuremath {\begin{array}[H]{c}
 {\scriptstyle { #1}} \\[-1mm] {=\!\!\!=\!\!\!=\!\!\!} \end{array} }}

\def \pso {p\hspace{-0.43em}/ _1}
\def \pst {p\hspace{-0.43em}/ _2}
\def \ksp {k\hspace{-0.49em}/_+}
\def \ksm {k\hspace{-0.49em}/_-}
\def \kso {k\hspace{-0.49em}/_1}
\def \kst {k\hspace{-0.49em}/_2}

\def \uo{\ensuremath { u(p_1) \, }}
\def \ut{\ensuremath { \bar{u}(-p_2) \, }}
\def \vo{\ensuremath { v(-p_3) \, }}
\def \vt{\ensuremath { \bar{v}(p_4) \, }}

\def\m{\ensuremath} 

\def \co#1{\ensuremath{ [ (k - q_{#1})^2 ]}}
\def \c#1{\ensuremath{ [ (k - q_{#1})^2 - m_{#1}^2 ]}}
\def \ck#1{\ensuremath{ [ k^2 - m_{#1}^2 ]}}

\def \FP#1#2{\ensuremath { \frac{ i}{[\,{#1} ({#2})-m+i\,\eps\,]} \,\,}}
\def \PP#1#2#3{\ensuremath { \frac{(- i \h{0.5})\,  \G_{#1 #2}}{[\,{#3}^2+i\eps\,]} \,\,}}
\def \BP#1#2#3#4{\ensuremath {\,\frac{ (-i\,e) \,  \G_{#1 #2}}{[\,{#3}^2-{#4}^2+i\,\eps\,]} \,\, }}

\def \Ve#1{\ensuremath { (-i\,e \, Q_e \, \ga^{#1} ) \,\, }}
\def \Vt#1{\ensuremath { (-i\,e \, Q_t \,  \ga^{#1} ) \,\, }}

\def \Vee#1{\ensuremath { (i\,e \, v_e \, \ga^{#1} ) \,\, }}
\def \Vtt#1{\ensuremath { (i\,e \, v_t \,  \ga^{#1} ) \,\, }}

\def \Vze#1{\ensuremath { \big(i\,e \, (v_e + a_e\, \ga_5)\,
\ga^{#1} \big) \,\, }}
\def \Vzt#1{\ensuremath { \big(i\,e \, (v_t + a_t\, \ga_5)\, \ga^{#1} \big) \,\, }}
\def \Vzeo#1{\ensuremath { \big(i\,e \, (v_e^0 + a_e^0\, \ga_5)\,
\ga^{#1} \big) \,\, }}
\def \Vzto#1{\ensuremath { \big(i\,e \, (v_t^0 + a_t^0\, \ga_5)\,
\ga^{#1} \big) \,\, }}

\def \Vzere#1{\ensuremath { \big(i\,e \, (\tilde{v_e})\,
\ga^{#1} \big) \,\, }}
\def \Vztre#1{\ensuremath { \big(i\,e \, (\tilde{v_t})\,
\ga^{#1} \big) \,\, }}

\def \Vw#1{\ensuremath { \left( \frac{i\,e }{ \sqrt{2} \, \sw} \, \ga^{#1} \right) \,\, }}

\def \BD#1#2{\ensuremath {\frac{(-i)}{[\,{#1}^2-{#2}^2+i\,\eps\,]} \,\, }}


\def \Vpww#1#2#3#4#5#6#7{\ensuremath { ({#1} i\,e)\, \Big[
    (-{#6}+{#7})^{\,#2} \G^{\,{#3}\,{#4}} -   ({#7}+{#5})^{\,#3} \,
    \G^{\,{#2}\, {#4}} + ({#5}+{#6})^{\,#4} \, \G^{\,{#2} \, {#3}} \Big]\,\,  }}

\def \Vzww#1#2#3#4#5#6#7{\ensuremath { \left(\frac{{#1} -  i\,e \, \cw}{\sw}\right)\, \Big[
    (-{#6}+{#7})^{\,#2} \G^{\,{#3}\,{#4}} -   ({#7}+{#5})^{\,#3} \,
    \G^{\,{#2}\, {#4}} +  ({#5}+{#6})^{\,#4} \, \G^{\,{#2} \, {#3}} \Big]\,\,  }}

\def \e#1#2{\ensuremath { \eps_{\,#1}^{\,*}\,({#2})\,\, }}

\def \BPCu#1#2#3#4{\ensuremath {\frac{(-i)}{[\,{#3}^2-{#4}^2+i\,\eps\,]}\,
    \left( \, \G^{\, {#1}\, {#2}} + \frac{{#3}_{#1}\,
        {#3}_{#2}}{\vec{#3}^2} - \frac{{#3}_0\, [{#3}_{#1}\,
        n_{#2}+{#3}_{#2}\,n_{#1}\, ]}{\vec{#3}^2} \right)  \,\, }}

\def \eett {\ensuremath {e^+e^- \to t \bar{t} \,\,}}
\def \GeV {\ensuremath  \,{\rm GeV} \,\,}

\def\sig{\left[\frac{\displaystyle{\mathrm{d}\sigma}}{\displaystyle{\mathrm{d}\cos \, \theta}}\right]}

\def\unity{{\rm 1\mskip-4.25mu l}}
\def\re{\mathop{\mathrm{Re}}}

\newcommand{\eqir}{\stackrel{{ \rm IR }}{\longrightarrow}}


%% file: p-v2.bbl
\begingroup\begin{thebibliography}{10}

\bibitem{Bhabha:1936xx}
H.~Bhabha, {\em Proc. Roy. Soc.} {\bf A154} (1936) 195.

\bibitem{Consoli:1979xw}
M.~Consoli, {\em Nucl. Phys.} {\bf B160} (1979)
208.

\bibitem{Bohm:1984yt}
M.~B{\"o}hm, A.~Denner, W.~Hollik, and R.~Sommer, {\em Phys. Lett.} {\bf B144}
  (1984)
414.

\bibitem{Tobimatsu:1985pp}
K.~Tobimatsu and Y.~Shimizu, {\em Prog. Theor. Phys.} {\bf 75} (1986)
905.

\bibitem{Bohm:1986fg}
M.~B{\"o}hm, A.~Denner, and W.~Hollik, {\em Nucl. Phys.} {\bf B304} (1988)
687.

\bibitem{Kuroda:1987yi}
S.~Kuroda, T.~Kamitani, K.~Tobimatsu, S.~Kawabata, and Y.~Shimizu, {\em Comput.
  Phys. Commun.} {\bf 48} (1988)
335--351.

\bibitem{Bardin:1991xe}
D.~Bardin, W.~Hollik, and T.~Riemann, {\em Z. Phys.} {\bf C49} (1991)
485--490.

\bibitem{Beenakker:1991mb}
W.~Beenakker, F.~Berends, and S.~van~der Marck, {\em Nucl. Phys.} {\bf B349}
  (1991)
323--368.

\bibitem{Montagna:1993py}
G.~Montagna, F.~Piccinini, O.~Nicrosini, G.~Passarino, and R.~Pittau, {\em
  Nucl. Phys.} {\bf B401} (1993)
3--66.

\bibitem{Field:1995dk}
J.~Field and T.~Riemann, {\em Comput. Phys. Commun.} {\bf 94} (1996) 53--87,
\href{http://www.arXiv.org/abs/hep-ph/9507401}{hep-ph/9507401}.

\bibitem{Beenakker:1998fi}
W.~Beenakker and G.~Passarino, {\em Phys. Lett.} {\bf B425} (1998) 199--207,
\href{http://www.arXiv.org/abs/hep-ph/9710376}{hep-ph/9710376}.

\bibitem{Kublbeck:1990xc}
J.~K{\"u}blbeck, M.~B{\"o}hm, and A.~Denner, {\em Comput. Phys. Commun.} {\bf
  60} (1990)
165--180.

\bibitem{Hahn:2000kx}
T.~Hahn, {\em Comput. Phys. Commun.} {\bf 140} (2001) 418--431,
\href{http://arXiv.org/abs/hep-ph/0012260}{hep-ph/0012260}.

\bibitem{Belanger:2003sd}
G.~Belanger {\em et al.},
\href{http://www.arXiv.org/abs/hep-ph/0308080}{hep-ph/0308080}.

\bibitem{Lorca:2004fg}
A.~Lorca and T.~Riemann, {\em Comput. Phys. Commun.} {\bf 174} (2006) 71--82,
\href{http://www.arXiv.org/abs/hep-ph/0412047}{hep-ph/0412047}.

\bibitem{Jadach:1996is}
S.~Jadach, W.~Placzek, E.~Richter-Was, B.~Ward, and Z.~Was, {\em Comput. Phys.
  Commun.} {\bf 102} (1997)
229--251.

\bibitem{Melles:1997qa}
M.~Melles, {\em Acta Phys. Polon.} {\bf B28} (1997) 1159--1206,
\href{http://arXiv.org/abs/hep-ph/9612348}{hep-ph/9612348}.

\bibitem{Arbuzov:1995qd}
A.~Arbuzov {\em et al.}, {\em Nucl. Phys.} {\bf B485} (1997) 457--502,
\href{http://www.arXiv.org/abs/hep-ph/9512344}{hep-ph/9512344}.

\bibitem{Arbuzov:1996jj}
A.~Arbuzov {\em et al.}, {\em Nucl. Phys. Proc. Suppl.} {\bf 51C} (1996)
  154--163,
\href{http://arXiv.org/abs/hep-ph/9607228}{hep-ph/9607228}.

\bibitem{Arbuzov:1999db}
A.~Arbuzov,
\href{http://arXiv.org/abs/hep-ph/9907298}{hep-ph/9907298}.

\bibitem{Kobel:2000aw}
{Two Fermion Working Group} Collaboration, M.~Kobel {\em et al.},
\href{http://www.arXiv.org/abs/hep-ph/0007180}{hep-ph/0007180}.

\bibitem{Jadach:2003zr}
S.~Jadach,
\href{http://www.arXiv.org/abs/hep-ph/0306083}{hep-ph/0306083}.

\bibitem{Aguilar-Saavedra:2001rg}
{ECFA/DESY LC Physics Working Group} Collaboration, J.~Aguilar-Saavedra {\em et
  al.}, DESY 2001--01 (2001),
\href{http://www.arXiv.org/abs/hep-ph/0106315}{hep-ph/0106315}.

\bibitem{Bern:2000ie}
Z.~Bern, L.~Dixon, and A.~Ghinculov, {\em Phys. Rev.} {\bf D63} (2001) 053007,
\href{http://www.arXiv.org/abs/hep-ph/0010075}{hep-ph/0010075}.

\bibitem{Bonciani:2004gi}
R.~Bonciani, A.~Ferroglia, P.~Mastrolia, E.~Remiddi, and J.~van~der Bij, {\em
  Nucl. Phys.} {\bf B701} (2004) 121--179,
\href{http://www.arXiv.org/abs/hep-ph/0405275}{hep-ph/0405275}.

\bibitem{Czakon:2004tg}
M.~Czakon, J.~Gluza and T.~Riemann,
{\em Nucl.\ Phys.\ Proc.\ Suppl.}  {\bf 135} (2004) 83,
\href{http://www.arXiv.org/abs/hep-ph/0406203}{hep-ph/0406203}.


\bibitem{Czakon:2004wm}
M.~Czakon, J.~Gluza, and T.~Riemann, {\em Phys. Rev.} {\bf D71} (2005) 073009,
\href{http://www.arXiv.org/abs/hep-ph/0412164}{hep-ph/0412164}.

\bibitem{Penin:2005kf}
A.~Penin, {\em Phys. Rev. Lett.} {\bf 95} (2005) 010408,
\href{http://www.arXiv.org/abs/hep-ph/0501120}{hep-ph/0501120}.

\bibitem{Bonciani:2005im}
R.~Bonciani and A.~Ferroglia,
{\em Phys.\ Rev.}\  {\bf D72} (2005) 056004,
\href{http://www.arXiv.org/abs/hep-ph/0507047}{hep-ph/0507047}.

%

\bibitem{Korner:2004rr}
J.~K{\"o}rner, Z.~Merebashvili, and M.~Rogal, {\em Phys. Rev.} {\bf D71} (2005)
  054028,
\href{http://www.arXiv.org/abs/hep-ph/0412088}{hep-ph/0412088}.

\bibitem{Korner:2005rg}
J.~K{\"o}rner, Z.~Merebashvili, and M.~Rogal, {\em Phys. Rev.} {\bf D73} (2006)
  034030,
\href{http://www.arXiv.org/abs/hep-ph/0511264}{hep-ph/0511264}.

\bibitem{Fleischer:2004ah}
J.~Fleischer, A.~Lorca, and T.~Riemann,
\href{http://www.arXiv.org/abs/hep-ph/0409034}{hep-ph/0409034}.

\bibitem{Lorca:2004dk}
A.~Lorca and T.~Riemann, {\em Nucl. Phys. Proc. Suppl.} {\bf 135} (2004) 328,
\href{http://www.arXiv.org/abs/hep-ph/0407149}{hep-ph/0407149}.

\bibitem{Gluza:2004tq}
J.~Gluza, A.~Lorca, and T.~Riemann, {\em Nucl. Instrum. Meth.} {\bf 534} (2004)
  289,
\href{http://www.arXiv.org/abs/hep-ph/0409011}{hep-ph/0409011}.

\bibitem{Lorca:2005yp}
A.~Lorca, DESY-THESIS-2005-004 (2005).

\bibitem{Tentyukov:1999is}
M.~Tentyukov and J.~Fleischer, {\em Comput. Phys. Commun.} {\bf 132} (2000)
  124--141,
\href{http://arXiv.org/abs/hep-ph/9904258}{hep-ph/9904258}.

\bibitem{Nogueira:1993ex}
P.~Nogueira, {\em J. Comput. Phys.} {\bf 105} (1993) 279.

\bibitem{Vermaseren:2000nd}
J.~Vermaseren,
\href{http://www.arXiv.org/abs/math-ph/0010025}{math-ph/0010025}.

\bibitem{Hahn:1998yk}
T.~Hahn and M.~P{\'e}rez-Victoria, {\em Comput. Phys. Commun.} {\bf 118} (1999)
  153,
\href{http://arXiv.org/abs/hep-ph/9807565}{hep-ph/9807565}.

\bibitem{vanOldenborgh:1991yc}
G.~J. van Oldenborgh, {\em Comput. Phys. Commun.} {\bf 66} (1991)
1.

\bibitem{Beenakker:1990jr}
W.~Beenakker and A.~Denner, {\em Nucl. Phys.} {\bf B338} (1990)
349--370.

\bibitem{Fleischer:2003kk}
J.~Fleischer, A.~Leike, T.~Riemann, and A.~Werthenbach, {\em Eur. Phys. J.}
  {\bf C31} (2003) 37,
\href{http://www.arXiv.org/abs/hep-ph/0302259}{hep-ph/0302259}.

\bibitem{webbhabha1}
T.~Riemann {\em et al.}, http://www-zeuthen.\linebreak[2]desy.\linebreak[2]de/%
  \linebreak[2]theory/\linebreak[2]research\linebreak[2]/bhabha/bhabha1.html.

\bibitem{web-masters:2004nn}
M.~Czakon, J.~Gluza, and T.~Riemann,
  http://www-zeuthen.desy.de/\-theory/\-research/\linebreak[2]bha\linebreak[2]%
bha\linebreak[2]/bha\linebreak[2]bha.html.

\bibitem{'tHooft:1972fi}
G.~'t~Hooft and M.~Veltman, {\em Nucl. Phys.} {\bf B44} (1972)
189--213.

\bibitem{Davydychev:2000na}
A.~Davydychev and M.~Kalmykov, {\em Nucl. Phys.} {\bf B605} (2001) 266--318,
\href{http://arXiv.org/abs/hep-th/0012189}{hep-th/0012189}.

\bibitem{Davydychev:2003mv}
A.~Davydychev and M.~Kalmykov,
{\em Nucl.\ Phys.}  {\bf B699} (2004) 3,
\href{http://arXiv.org/abs/hep-th/0303162}{hep-th/0303162}.


\bibitem{Kalmykov:2006pu}
M. Kalmykov,
{\em JHEP} {\bf 0604} (2006) 056,
\href{http://arXiv.org/abs/hep-th/0602028}{hep-th/0602028}.

\bibitem{webpage-kalmykov}
M.~Kalmykov,
  http://theor.jinr.ru/{${\sim}$}kalmykov/.


\bibitem{Kotikov:1991hm}
A.~Kotikov, {\em Phys. Lett.} {\bf B259} (1991)
314--322.

\bibitem{Remiddi:1997ny}
E.~Remiddi, {\em Nuovo Cim.} {\bf A110} (1997) 1435--1452,
\href{http://www.arXiv.org/abs/hep-th/9711188}{hep-th/9711188}.

\bibitem{Remiddi:1999ew}
E.~Remiddi and J.~Vermaseren, {\em Int. J. Mod. Phys.} {\bf A15} (2000)
  725--754,
\href{http://www.arXiv.org/abs/hep-ph/9905237}{hep-ph/9905237}.

\bibitem{looptools}
T.~Hahn, The LoopTools Homepage, http://www.feynarts.de/looptools/.

\bibitem{Fleischer:2003rm}
J.~Fleischer, F.~Jegerlehner, and O.~Tarasov, {\em Nucl. Phys.} {\bf B672}
  (2003) 303--328,
\href{http://www.arXiv.org/abs/hep-ph/0307113}{hep-ph/0307113}.

\bibitem{Fleischer:2003bg}
J.~Fleischer, T.~Riemann, and O.~Tarasov, {\em Acta Phys. Polon.} {\bf B34}
  (2003)
5345--5356.

\bibitem{Laporta:2000dc}
S.~Laporta, {\em Phys. Lett.} {\bf B504} (2001) 188--194,
\href{http://arXiv.org/abs/hep-ph/0102032}{hep-ph/0102032}.

\bibitem{Laporta:2001dd}
S.~Laporta, {\em Int. J. Mod. Phys.} {\bf A15} (2000) 5087--5159,
\href{http://arXiv.org/abs/hep-ph/0102033}{hep-ph/0102033}.

\bibitem{Laporta:2003jz}
S.~Laporta,
\href{http://www.arXiv.org/abs/hep-ph/0311065}{hep-ph/0311065}.

\bibitem{mathworld}
{Wolfram Research}, http://mathworld.wolfram.com.

\bibitem{Prudnikov:1986III}
A.~Prudnikov, Y.~Brychkov, and O.~Marichev, {\em Integrals and Series, Vol. 3:
  More Special Functions}.
\newblock Nauka, Moskva, 1986.

\bibitem{Bonciani:2003cj}
R.~Bonciani, A.~Ferroglia, P.~Mastrolia, E.~Remiddi, and J.~van~der Bij, {\em
  Nucl. Phys.} {\bf B681} (2004) 261--291,
\href{http://www.arXiv.org/abs/hep-ph/0310333}{hep-ph/0310333}.

\bibitem{Gehrmann:2000zt}
T.~Gehrmann and E.~Remiddi, {\em Nucl. Phys.} {\bf B601} (2001) 248--286,
\href{http://www.arXiv.org/abs/hep-ph/0008287}{hep-ph/0008287}.

\bibitem{Smirnov:2004}
V.~Smirnov, {``Evaluating Feynman Integrals''} (Springer Verlag, Berlin, 2004).

\bibitem{Czakon:2005rk}
M.~Czakon,
\href{http://www.arXiv.org/abs/hep-ph/0511200}{hep-ph/0511200}.

\bibitem{Moch:2005uc}
S.~Moch and P.~Uwer, {\em Comput. Phys. Commun.} {\bf 174} (2006) 759--770,
\href{http://www.arXiv.org/abs/math-ph/0508008}{math-ph/0508008}.

\bibitem{Davydychev:1991va}
A.~Davydychev, {\em Phys. Lett.} {\bf B263} (1991)
107--111.

\bibitem{Tarasov:1996br}
O.~Tarasov, {\em Phys. Rev.} {\bf D54} (1996) 6479--6490,
\href{http://arXiv.org/abs/hep-th/9606018}{hep-th/9606018}.

\bibitem{Fleischer:1999hq}
J.~Fleischer, F.~Jegerlehner, and O.~V. Tarasov, {\em Nucl. Phys.} {\bf B566}
  (2000) 423--440,
\href{http://arXiv.org/abs/hep-ph/9907327}{hep-ph/9907327}.

\bibitem{Tkachov:1981wb}
F.~Tkachov, {\em Phys. Lett.} {\bf B100} (1981)
65--68.

\bibitem{Chetyrkin:1981qh}
K.~Chetyrkin and F.~Tkachov, {\em Nucl. Phys.} {\bf B192} (1981)
159--204.

\end{thebibliography}\endgroup
